# Decentralized Finance (Literacy) today and in 2034: Initial Insights from Singapore and beyond

Daniel Liebau*
First version: Dec, 2024. Updated: June 2025

## Abstract

How will Decentralized Finance transform financial services? Using New Institutional Economics and Dynamic Capabilities Theory, I analyse survey data from 109 experts using non-parametric methods. Experts span traditional finance, DeFi industry, and academia. Four insights emerge: adoption expectations rise from negligible to 43% expecting at least high adoption by 2034; experts expect convergence scenarios over disruption, with traditional finance embracing DeFi most likely; back-office transforms before customer-facing functions; strategic competencies eclipse DeFi-sector specific- and technical skills. This challenges technology-centric adoption models. DeFi represents emerging market entry requiring organizational transformation, not just technological implementation. SEC developments validate predictions. Financial institutions should prioritize developing strategic capabilities over mere technical training.

---

* Author: Dan Liebau is an Affiliate Faculty member for Digital Assets, Blockchain and Decentralized Finance at Singapore Management University, a Lecturer for FinTech at ESSEC Business School, and a Lecturer and PhD Candidate with the Finance department of Rotterdam School of Management, Erasmus University, PO Box 1738, 3000 DR Rotterdam, The Netherlands, liebau@rsm.nl.

I am grateful to: Nuno Almeida Camacho, Thomas Lambert and Peter Roosenboom, all at Erasmus University, Simon Trimborn at University of Amsterdam, Cindy Deng at Nanyang Technological University in Singapore, Patrick Schueffel, Angelo Aspris at University of Sydney, Lennart Ante at Constructor University and Sinclair Davidson at RMIT in Australia. Thanks to several practitioners for their feedback, too: Emma Channing of Satis, Roberto Durscki at Stellar, Lee Schneider at Ava Labs, and Greg Murphy at Kaimeta. I thank the team at SMU Academy, in particular Michael Low and Yew Keong Poon who kindly supported me with the data collection. I also want to thank organizers and participants of the 6th Blockchain International Scientific Conference (ISC2024) in Singapore for the opportunity to present and the valuable feedback I collected. Thanks as well for feedback to the FC2025 (FinTeAchIn track) conference participants. I also appreciate the opportunity to present my draft finding at the 24Fintech conference in Riyadh, Saudi Arabia and Istanbul FinTechWeek in Istanbul. I am solely responsible for the work and any errors or omissions.

## 1. Introduction

Financial institutions move slowly, but technology-based change accelerates. This temporal mismatch creates a dilemma for established firms facing the Decentralized Finance (DeFi) phenomenon, a blockchain-based eco-system providing financial services without traditional intermediaries. Developing organizational capabilities takes years, but waiting risks obsolescence. Malinova and Park (2024) theoretically demonstrate DeFi could reduce U.S. equity trading costs by up to USD 15 billion annually, highlighting its transformative potential. The question is not whether financial institutions should prepare, but how, and what capabilities will be critical to succeed in times ahead.

This paper addresses two interrelated questions shaping the future of finance. First, how will DeFi transform traditional financial services over the next decade? Second, what institutional competencies will organizations need to succeed in this transformed landscape? While recent research examines DeFi's technical architecture and immediate risks, the field lacks systematic assessment of long-term institutional competency requirements. This gap matters because capability development is far less costly than competitive displacement.

I examine these questions through expert elicitation, surveying 109 experts across three groups: Traditional Finance Professionals (including regulators), DeFi industry practitioners, and DeFi-focused Academic Researchers. The ten-year horizon, to 2034, provides sufficient time for institutional adaptation while remaining within reasonable forecasting bounds. Expert perceptions matter particularly regarding emerging phenomena where today's expectations shape tomorrow's strategic- and investment decisions, eventually creating self-fulfilling dynamics.

Using non-parametric statistical methods, I analyse responses through two theoretical lenses. New Institutional Economics (NIE) illuminates how DeFi creates alternative institutional arrangements and why certain business functions face greater transformation, while Dynamic Capabilities Theory (DCT) explains both the organizational adaptations required and why different expert groups perceive these changes differently. The dual theoretical lens allows me to analyse expert predictions about DeFi's future role and the competencies it will require from financial services professionals.

The study makes three main contributions. First, I provide the first comprehensive expert assessment of DeFi's expected impact on financial services by 2034, addressing a critical gap in forward-looking institutional analysis. Second, I introduce "DeFi literacy" as a multidimensional construct capturing organizational capabilities required for DeFi integration, extending financial literacy theory from individual to institutional contexts. Third, I document significant differences in DeFi expectations across expert groups and regions, revealing how proximity to the technology shapes perceptions of institutional transformation.

Four insights emerge with immediate relevance for financial institutions. First, adoption expectations diverge sharply today but converge for 2034: while no expert group rates current institutional adoption as at least high, 43% anticipate at least high adoption within a decade. Second, out of five scenarios presented, those suggesting convergence rather than competition appear as the most likely. Third, DeFi will reshape how finance operates before changing what it offers customers. Experts expect major impacts on risk management (86%) and operations (80%), but modest effects on customer service. Fourth, and most striking, strategic and organizational competencies will eclipse technical skills. The perceived importance of strategic capabilities rises from 60% today to 84% by 2034, surpassing both sector-specific DeFi knowledge and technological skills. This suggests, paradoxically, that DeFi literacy is less about understanding blockchain than mastering organizational transformation.

These insights reframe the DeFi debate from technology adoption to institutional evolution and emerging market entry. Understanding DeFi as an emerging market rather than merely a technology reveals why 'DeFi literacy' represents a fundamental shift. Like entering any emerging market, organizations must develop capabilities for navigating unfamiliar institutional arrangements, governance structures, and risk profiles. With 68% of experts citing capability gaps as critical barriers, institutions face a clear imperative: begin building multidimensional competencies now. Recent regulatory developments, including SEC statements through to June 2025 supporting DeFi's potential while emphasizing regulated integration, validate several expert predictions and reinforce practical relevance.

Following Graham and Harvey's (2001) approach to capturing practitioner perspectives, I hope insights will be helpful to finance scholars. At the same time, survey responses might also be valuable for policymakers, regulators, financial services executives and DeFi industry practitioners as well as learning & development leaders.

The paper is structured as follows. Section 2 documents the theoretical framework. Section 3 positions this study within the existing literature. Section 4 describes the expert elicitation methodology. Section 5 details the analytical approach. Section 6 presents comparative results across expert groups and regions. Section 7 concludes.

## 2. Theoretical framework

This study employs two complementary theoretical frameworks to analyse DeFi's potential impact on financial services and required competencies: New Institutional Economics and Dynamic Capabilities Theory. Together, these frameworks provide suitable lenses to understand both the industry-level institutional changes DeFi may trigger and the organizational capabilities financial institutions will need to develop in response.

*NIE and DeFi*

NIE developed through the work of scholars including North (1990), Williamson (1985), and Ostrom (2005), examines how institutional arrangements evolve in response to transaction costs, information asymmetries, and governance requirements. NIE is particularly relevant to understanding DeFi, which fundamentally reimagines financial institutions through technological means. Traditional financial institutions emerged as solutions to problems of information asymmetry, high transaction costs, and trust requirements as documented by Diamond and Dybvig (1983) and Leland and Pyle (1977). Banks and other intermediaries aim to reduce these costs by leveraging economies of scale, specialized knowledge, and regulatory guarantees. However, institutions also introduce agency costs, regulatory overhead, and potential inefficiencies, for example in the form of technology debt.

DeFi represents a novel institutional arrangement that addresses these same economic problems traditional finance aims to solve, but through different mechanisms: replacing human intermediation with algorithmic processes, trust in institutions with cryptographic verification, and centralized governance with distributed consensus mechanisms (Schär, 2021 and Harvey et al., 2021). Through the NIE lens, DeFi can be understood as an institutional innovation that might reduce certain transaction costs. But it most certainly also introduces new governance challenges. The degree to which DeFi will be adopted depends on its relative efficiency in addressing the fundamental economic problems that financial institutions solve. As Williamson (1979) argues, institutional arrangements that minimize transaction costs will eventually predominate. This perspective helps explain why DeFi might be adopted more readily in areas where traditional finance faces significant friction costs, such as cross-border transactions, while facing greater challenges in areas where existing institutions have already optimized efficiency. The institutional perspective also helps explain the likely evolution of DeFi's relationship with traditional finance. Drawing on the concept of institutional isomorphism (DiMaggio and Powell, 1983), it may be reasonable to expect regulatory pressures to drive increasing similarity between DeFi and traditional financial services over time.

*Dynamic Capabilities and DeFi Literacy*

While NIE helps understand industry-level evolution, DCT provides a framework for analysing how individual organizations can adapt to rapid change. Introduced by Teece, Pisano, and Shuen (1997) and further developed by Teece (2007), this theoretical framework examines how firms develop capabilities to "integrate, build, and reconfigure internal and external competences to address rapidly changing environments". In the context of DeFi, financial institutions with strong dynamic capabilities will be better positioned to thrive amid the institutional changes that DeFi may trigger. Their categorization of dynamic capabilities into sensing, seizing, and transforming provides the framework for understanding different aspects of DeFi literacy: Sensing capabilities include understanding DeFi technology, monitoring ecosystem developments, and identifying potential opportunities or threats. Seizing capabilities involve designing appropriate responses to DeFi developments, making strategic decisions about adoption or competition, and allocating resources effectively. Transforming capabilities encompass implementing organizational changes, developing new competencies, and reconfiguring business models in response to DeFi's evolution.

The dynamic capabilities perspective is particularly valuable for understanding the "DeFi literacy" that financial professionals and organizations will need. This extends traditional concepts of financial literacy (Lusardi and Mitchell, 2014) to encompass the more technical, strategic, and operational capabilities needed by institutions in a DeFi-influenced financial services environment of the future rather than focusing on an individual's competencies. As financial literacy focuses on knowledge and skills needed for effective financial decision-making, DeFi literacy encompasses the capabilities needed to operate effectively in an environment where decentralized and traditional finance coexist and eventually converge.

Recent empirical evidence validates the dynamic capabilities framework to assess DeFi adoption. Organizational blockchain adoption studies demonstrate that financial institutions developing dynamic capabilities show improved financial performance compared to non-adopters (Khalil et al, 2021). The functional approach to DeFi proposed by Aquilina et al. (2024) aligns with my framework, showing how institutions must develop capabilities across five functional layers: settlement, asset, protocol, application, and aggregation. This multi-layered capability requirement contributes to our understanding on why strategic competencies emerge as more critical than purely DeFi sector-specific and technical skills during the expert elicitation.

*Integrated Theoretical Approach*

Together, NIE and DCT provide complementary perspectives on DeFi's potential impact and capability requirements. NIE helps explain the macro-level institutional changes that may occur in the financial services industry, while DCT helps understand what capabilities organizations might need to respond to these changes. This integrated theoretical approach guides my interpretation of expert survey submissions. Experts' views on adoption levels, platform roles, and future scenarios can be understood as predictions about institutional evolution, while their assessments of business area impacts and required competencies reflect judgments about necessary dynamic capabilities.

## 3. Literature review

Next, I review relevant literature, organized through the theoretical lenses of NIE and DCT.

*Institutional Arrangements and Governance*

DeFi develops and morphs swiftly, and so does the related literature. In their early account, Harvey et al. (2021) frame DeFi's value proposition in terms that align with NIE principles, identifying five problems that DeFi could help solve: centralized control, limited access, inefficiency, lack of interoperability, and opacity. These problems represent institutional shortcomings that DeFi attempts to address through alternative governance mechanisms. Alongside these opportunities, authors also describe idiosyncratic risks of DeFi: smart-contract risk, governance risk, oracle risk, scaling risk, impermanent loss risk and regulatory risk. Table 1 provides definitions of these opportunities and risks. For financial services professionals, understanding both is essential to manage the future potential impact of DeFi.

**Table 1**. DeFi Opportunities and Risks. Note: Unless otherwise indicated definitions are deductions from Harvey et al. (2021).

**Panel A**. DeFi Opportunities

| Opportunities | Definitions |
| --- | --- |
| Reduce centralized control | Increase (product) options for financial services users and reduce associated switching costs |
| Reduce limited access to finance | Enable access to affordable financial services for everyone |
| Reduce inefficiencies | The opportunity of reducing inefficiencies in financial services by leveraging digital/internet technology |
| Improve interoperability | Removing silos in financial services enabling swift processing across financial services |
| Reduce opacity | Increase transparency in the industry for users to better assess counterparty risks |

**Panel B**. DeFi Risks

| Risks | Definitions |
| --- | --- |
| Smart-contract risk | Risks arising from logical errors in code or economic exploits by rogue actors |
| Governance risk | The risk of a DAO's decision-making processes becoming dysfunctional* |
| Oracle risk | The risk of oracle services misfunctioning |
| Scaling risk | The risk of smart contract platforms not being able to timely process required transaction volume |
| Impermanent loss risk | The temporary token value loss faced by liquidity providers of automated market makers compared to holding tokens directly** |
| Regulatory risk | The potential negative effect of regulatory decisions on DeFi platforms and ecosystems |

* DAO Governance risk as per Liebau and Oh (2024)

** Impermanent loss as per Capponi and Jia (2021)

Buterin et al. (2024) investigate private DeFi transactions, addressing the institutional tension between transparency and confidentiality. Feichtinger et al. (2024) document governance-related risks in Decentralized Autonomous Organizations (DAOs), highlighting the challenges of creating effective governance institutions without traditional organizational hierarchies. Bongaerts et al. (2025) investigate vote delegation in DeFi Governance and show that reputation and merit matter when delegators allocate their votes to delegates.

*Transaction Costs and Market Structures*

An implicit NIE focus on transaction costs is evident in studies examining DeFi's economic efficiency. Malinova and Park (2024) provide theoretical evidence that applying DeFi concepts to US equity markets could reduce annual trading costs by USD 6.5-15 billion, with particularly significant implications for small firms and emerging markets. This aligns with NIE's prediction that institutional arrangements that reduce transaction costs will gain adoption.

Market structure studies include Rivera et al. (2023), who investigate equilibrium conditions in DeFi lending markets, and Capponi and Jia (2021) and Lehar and Parlour (2023), who examine pricing mechanisms and liquidity in decentralized exchanges. In a way, these studies analyse how DeFi creates new institutional arrangements for fundamental financial functions.

*Regulatory Frameworks and Institutional Isomorphism*

The regulatory dimension of NIE is addressed by several scholars. Makarov and Schoar (2022) identify tax compliance, anti-money laundering laws, and preventing financial malfeasance as key issues for DeFi, highlighting the importance of regulatory enforcement mechanisms. Zetsche et al. (2020) propose new regulatory approaches for DeFi, recognizing that institutional evolution requires adapted regulatory frameworks including embedding regulatory controls into smart contracts. The concept of institutional isomorphism is evidenced in Lim et al. (2023), who document experiments by the Monetary Authority of Singapore (MAS) and Bank for International Settlements (BIS) that explore how traditional regulatory concerns can be addressed within DeFi systems. Similarly, Bok (2024) introduces the concept of "institutional DeFi," which represents an isomorphic adaptation where traditional financial institutions adopt DeFi applications.

*DeFi Through the Dynamic Capabilities Lens*

The literature includes works that help organizations develop sensing capabilities for DeFi opportunities and threats. Auer et al. (2023) introduce the DeFi Stack Reference Model, providing a framework for understanding DeFi's layered architecture. This model helps organizations sense how different components of the DeFi ecosystem interact and identify potential opportunities.

John et al. (2023) describe the functioning of smart contracts and related use cases in DeFi, contributing to the knowledge base organizations need for effective sensing capabilities. Similarly, Lambert et al. (2021) and Kreppmeier et al. (2023) document security token offerings as early attempts to bridge traditional and decentralized finance.

The literature on risk management contributes to understanding what capabilities organizations need to seize opportunities and transform in response to DeFi. Catalini et al. (2021) examine stablecoins and identify unregulated algorithmic stablecoins as posing significant risks, highlighting the importance of risk assessment capabilities. Kassoul et al. (2024) investigate contagion risk on decentralized lending platforms, informing the systemic risk management capabilities organizations need to develop. The technical capabilities dimension is addressed by studies like Lim et al. (2023), who also discuss privacy-preserving technologies such as zero-knowledge proofs, fully homomorphic encryption, and multiparty computation as potential solutions to privacy challenges on public blockchains. These articles help financial services organizations understand what technical capabilities they need to develop to participate effectively in the DeFi ecosystem.

*Synthesis and Research Gaps*

While the existing literature provides valuable insights into specific aspects of DeFi, there remains a significant gap in understanding how DeFi might reshape the financial services industry over the longer term and what competencies financial professionals will need. Most current studies focus on mechanisms, short-term market behaviours, or specific use cases and governance. My study aims to address this gap by eliciting expert opinions. By using established theoretical frameworks to interpret expert forecasts, I contribute to a increasingly structured understanding of how DeFi might evolve as an institutional arrangement and what capabilities financial services organizations will need to develop in response.

*Financial Literacy and DeFi Literacy*

Financial literacy research has established the importance of individual competencies for financial decision-making (Lusardi and Mitchell, 2014 and Fernandes et al., 2014). Traditional financial literacy encompasses three core dimensions: numeracy (compound interest calculations), inflation comprehension, and risk diversification understanding (Lusardi and Mitchell, 2011). These individual-level competencies prove essential for personal financial wellbeing. They are not designed to assist with understanding organizational responses in financial services to rapid industry developments and the emergence of new market segments. I extend their framework to introduce "DeFi literacy" as organizational-level competencies required for navigating decentralized finance. Dong and Blankson (2025) propose a progressive learning model showing how organizations advance from interest through knowledge to insights in fintech adoption. Their 'Fin + Tech' framework demonstrates that successful blockchain adoption requires both financial domain expertise and technological capabilities, a finding that directly supports my multidimensional DeFi literacy construct. These theoretical extensions parallel how organizational learning theory evolved from individual learning models (Argyris and Schön, 1978). Just as organizations require different capabilities than individuals for knowledge creation and retention (Nonaka, 1994), DeFi literacy also demands institutional competencies specific to blockchain technology and cryptocurrencies, in addition to strategic competencies. In Jones et al. (2024) authors develop the Cryptocurrency Literacy Scale, identifying five domains of digital financial capability: achievement, tracking, planning, product selection, and being informed. Additionally, empirical studies show that organizational factors outweigh technological considerations in blockchain adoption (Saheb and Mamaghani, 2021), suggesting capability development is most critical.

Table 2 summarizes my contribution to advancing financial literacy theory by: shifting the level of analysis from individual to organizational, expanding beyond cognitive skills to include strategic and transformational capabilities, and incorporating technological and governance dimensions absent from traditional financial literacy. While financial literacy focuses on understanding existing financial primitives, DeFi literacy encompasses capabilities for navigating institutional transformation in finance.

**Table 2.** From Financial Literacy to DeFi Literacy

| Dimension | Financial Literacy (Individual) | DeFi Literacy (Organizational) |
|---|---|---|
| Level | Individual decision-making | Institutional capabilities |
| Core Knowledge | Interest rates, inflation, compound returns | Smart contracts, consensus mechanisms, DeFi primitives |
| Risk Understanding | Portfolio diversification, market risk | Smart contract risk, composability risk, governance risk |
| Key Skills | Numeracy, financial planning | Strategic vision, technical integration, DAO governance participation |
| Primary Focus | Personal wealth management | Organizational transformation |

# 3  The Survey

An ongoing study[1] that investigates the effects of AI on human capital in the future influences my work. This research project is also forward-looking and relies on expert elicitation via an online

---

[1] I refer here to the "AI Competences 2035 project" led by Prof Ahmed Bounfour. More information can be found on the project's website: http://www.chairedelimmateriel.universite-paris-saclay.fr/2023/07/03/ai-competences-2035/ (last accessed 8 Nov 2023)

survey. My survey documents expert's answers to the following questions: What is the perceived adoption of DeFi today, and what will it be in 2034? To what extent will DeFi change traditional finance? What do experts perceive DeFi's future role to be? What impact will DeFi have on business areas, processes, and customer service? What issues and problems may arise with the rise of DeFi? Lastly, which competencies will be most critical for financial services professionals to succeed in a future with DeFi?

I now describe the survey in more detail starting with administrational matters. After a short welcome message, I state that the survey duration will be of approximately 15-20 minutes and communicate the number of questions, 23 in total. I communicate the objective of the survey as "understand and anticipate the evolution of organization-level competencies in relation to Decentralized Finance and its impact on the Financial Services industry". The study is forward-looking, with a time horizon of approximately ten years. I chose this time horizon because new technological phenomena in Finance, a heavily regulated industry, are not broadly adopted within a short periods of time. On the other hand, other studies (Dion et al., 2020) use a 25-year time horizon, which might be too long in the context of fast-moving technology adoption. I leverage Auer's et al. (2023) definition of Decentralized Finance - "a new financial paradigm that leverages distributed ledger technologies to offer services such as lending, investing, or exchanging crypto assets without relying on a traditional centralized intermediary". I thank participants for taking the time to participate and highlight the importance of the project.

In the first section of the survey, I gather information about experts, including to which of the three groups they belong (DeFi Course Participants, DeFi Industry Practitioners or DeFi-focused Academic Researchers). In the main section of the survey, I elicit opinions about DeFi adoption today and in 2034, DeFi competencies, DeFi-related issues and potential problems, the impact of DeFi on the industry, the impact on business areas and processes and different possible scenarios for the year of 2034. I conclude the elicitation with a section asking for additional information about professional experience and location and ask for the participant's email address before again thanking them for participation. Table 3 details the online survey design.

**Table 3**. Overview of DeFi 2034 online survey.

| Section | Focus | Data type | Collection method | Scale |
|---|---|---|---|---|
| Introduction | n/a | n/a | n/a | n/a |
| Sample description 1 | n/a | Nominal | Multiple choice / free text | n/a |
| Level of DeFi adoption | Adoption | Ordinal | 5-point Likert scale | Unsure (0)-Very High (4) |
| Key DeFi competencies for financial services | Competencies | Ordinal | 5-point Likert scale | Unsure (0) – Very Important (4) |
| DeFi-related issues and potential problems | Adoption | Ordinal | 5-point Likert scale | Unsure – Very Important |
| DeFi Platform Role | Adoption | Ordinal | 6-point Likert scale | Unsure (0) – DeFi Dominance (5) |
| DeFi impact on business areas and processes | Adoption | Ordinal | 5-point Likert scale | Unsure (0) – High(4) |
| DeFi and Customer Service | Adoption | Ordinal | 5-point Likert scale | Unsure (0) – High(4) |
| Scenarios for 2034 | Adoption | Ordinal | 5-point Likert scale | Unsure (0) – Most Likely (4) |
| Sample description 2 | n/a | Nominal | Multiple choice / free text | n/a |
| Conclusion / Thank you note | n/a | n/a | n/a | n/a |

I implemented the questionnaire in google forms[2] and shared it with two DeFi experts[3] to establish content validity ahead of use. The next section describes the data collected using the online survey.

## 4 The Data

I collect survey responses between 8.8.2023 and 7.11.2023. Experts are part of either one of the three expert groups. First, DeFi Course Participants: To elicit expert opinions from the first group, I collaborate with the Singapore Management University (SMU) Academy's management team to send 353 invitation letters to all participants who took my 2-day executive education course titled "Decentralised Finance (DeFi): A New Financial Ecosystem" in the period from 14.10.2021 to 10.05.2023. Experienced traditional financial services professionals, including many officers from the Monetary Authority of Singapore, mainly attended the course, making this group an adequate representation of conservative finance professionals and regulators. I record a total of 50 responses, which equates to a response rate of ca. 14.2%. Second, a group of DeFi Industry Practitioners: I use my global LinkedIn network to collect survey responses from relevant DeFi Industry Practitioners. To identify suitable experts, I filter for my first-grade connections (ca. 21,000 at the time of writing in November of 2023) and search for the keywords "DeFi", which results in a total of 221 individuals. I also search for "Decentralised Finance", which results in 33 individuals. I then review relevant profiles individually. I exclude most marketing and sales profiles as they generally have a less detailed understanding pertinent to this study. Subsequently I send out 61 invitations to participate in the survey and receive 27 responses, equating to a response rate of 44.3%. I also receive 16 responses marked as "Other". I review their responses to a separate question about the respondents' professional backgrounds and deduce that all 16 are also DeFi Industry Practitioners. This increases responses to 77 in total and 43 for the DeFi Industry Practitioner group (55.8%). Third, the last group: DeFi-focused Academic Researchers: I also use my global LinkedIn network to identify academic experts. I filter for my first-grade connections and search for the keywords "PhD" and "Professor". I then review relevant profiles individually to ensure their relevance. I send out 35 invitations to participate in the survey and receive 16 responses, which equates to a response rate of ca. 45.7%. Overall, the survey had a total response rate of 23.4%, with 109 participants completing the online survey. Bojke et al. (2021) suggest that 20 experts are needed to make a meaningful contribution. Table 4 summarizes online survey responses.

---

[2] The survey instrument can be inspected here: https://forms.gle/Kg9iDPX79CuQpxJTy6
3 I thank: Sandy Oh, a fellow Affiliate Faculty at Singapore Management University and Kenneth Bok, founder of Blocks.sg and author of "Decentralizing Finance" (Wiley, 2023) for their assistance with testing and improving the online survey.

**Table 4.** Online survey responses. There are three expert groups: 1) The course participants of Singapore Management University's executive course titled "Decentralised Finance: A new financial ecosystem", 2) Handpicked DeFi Industry Practitioners with relevant working experience and 3) DeFi-focused Academic researchers.

|  | DeFi Course Participants | DeFi Industry Practitioners | DeFi-focused Academic Researchers | Total |
|---|---|---|---|---|
| Reached out to | 353 | 77 | 35 | 465 |
| Responded | 50 | 43 | 16 | 109 |
| Response rate | 14.2% | 55.8% | 45.7% | 23.4% |

Using the expert answers, I can distinguish respondents based on their group and industry experience to further characterise the sample, as shown in Figure 1. Panel A illustrates the distribution of participants by expert group. Most respondents were DeFi Course Participants, representing 45.9% of the sample. DeFi Industry Practitioners comprised 39.4% of the sample, while DeFi-focused Academic Researchers comprised 14.7%. Panel B displays the split of participants by industry experience. The largest group had 10-20 years of experience (36.7%), followed by those with over 20 years (23.9%). Participants with 1-5 years and 5-10 years of experience represented 22.9% and 16.5%, respectively.

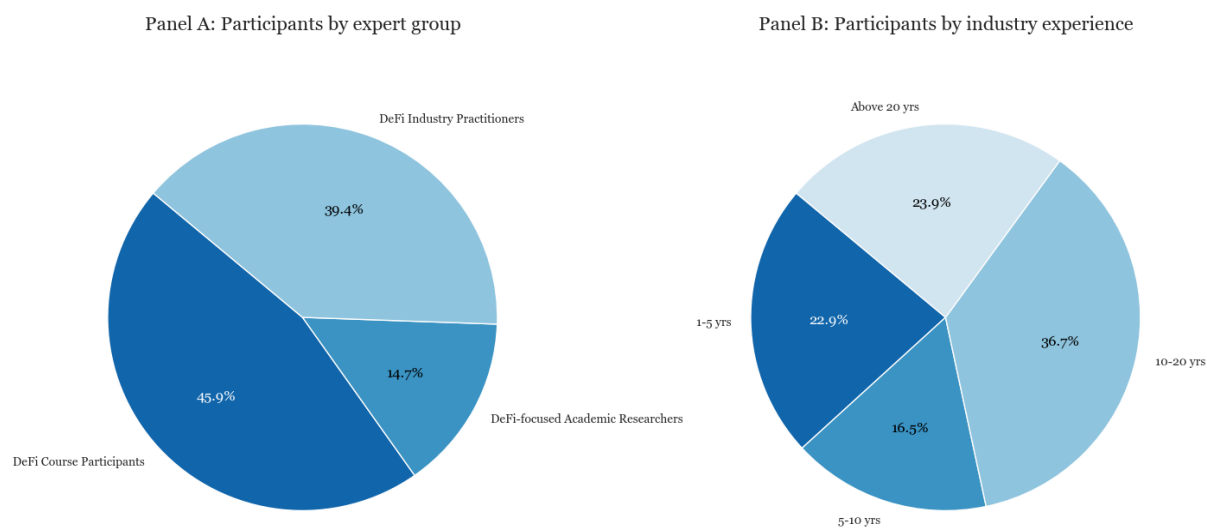

**Figure 1**. Participants by expert group and industry experience. There are three expert groups: 1) The DeFi Course Participants of Singapore Management University's executive course titled "Decentralised Finance: A new financial ecosystem", consisting mostly of senior traditional financial services professionals, 2) Handpicked DeFi Industry Practitioners with relevant DeFi working experience and 3) Handpicked DeFi-focused Academic Researchers. The survey is based on 109 expert responses.

I also ask respondents where their employer is headquartered. Most respondents answered with Singapore, representing 63 out of 109 participants. The USA and Switzerland followed with 11 and 8 respondents, respectively. Other countries with notable representation include Germany (5), the United Kingdom (5), and Hong Kong SAR (3). Several other countries, including Australia, Canada, and France, had a single respondent each. In contrast, a small group categorized as "Other" accounted for four additional respondents. Table 5 provides an overview of the geographical distribution of survey respondents' headquarters.

**Table 5.** Survey respondent's headquarters.

| Country | Count |
| --- | --- |
| Singapore | 63 |
| USA | 11 |
| Switzerland | 8 |
| Germany | 5 |
| United Kingdom | 5 |
| Hong Kong SAR | 3 |
| Australia | 1 |
| Canada | 1 |
| Cayman Islands | 1 |
| Chile | 1 |
| Denmark | 1 |
| France | 1 |
| Ireland | 1 |
| Luxembourg | 1 |
| Netherlands | 1 |
| Poland | 1 |
| Other | 4 |
| **Total** | **109** |

Throughout this study, I refer to the three expert groups as Course Participants (CP) representing traditional finance professionals, Industry Practitioners (IP) representing DeFi industry practitioners, and Academic Researchers (AR) representing DeFi-focused academic researchers. All in all, the variation in experts' characteristics permits a sound description of how DeFi might develop.

## 5 Statistical Analysis Approach

This study primarily collected ordinal data through Likert-type scales. I adopt non-parametric approaches to analyse differences between groups on ordinal measures. For comparing responses between expert groups (DeFi Course Participants, DeFi Industry Practitioners, and DeFi-focused Academic Researchers) and experience levels (1-5 years, 5-10 years, 10-20 years, and 20+ years), I employ:

- The Kruskal-Wallis H test to determine statistically significant differences between two or more groups on ordinal dependent variables.
- The Mann-Whitney U test for post-hoc pairwise comparisons when the Kruskal-Wallis test indicates significant differences, with Bonferroni correction applied to adjust for multiple comparisons.
- Median and interquartile range (IQR) as descriptive measures instead of means when appropriate, to represent central tendency and dispersion in ordinal data.

This approach allows me to identify significant differences in perceptions between expert groups and experience levels and provides insights into how different experts view DeFi's potential impact and required competencies.

## 6 Insights

Insights presented in this section are based on the expert elicitation conducted via the survey and are interpreted through the dual lenses of NIE and DCT: NIE helps us understand expert predictions about DeFi's institutional evolution, while DCT also assists in understanding the organizational adaptations that may be required.

I start discussing current and 2034 adoption levels, then ask about the role of DeFi platforms, and possible future scenarios. Thereafter, I ask about impact on different business areas and DeFi-related issues and problems. Then, I move to DeFi competencies required in the future, what can now be called DeFi literacy. To make the article accessible to a wide audience, it differs from traditional research articles in that the common results and discussion sections are combined.

*Current and 2034 adoption levels*

To elicit expert opinions regarding DeFi adoption levels I ask, "How would you assess the current/2034 levels of DeFi adoption within the financial services industry as a whole?". Experts are asked to rate on a 5-point scale: 0 (Unsure), 1 (Low: DeFi will be used in limited contexts or only by few organizations), 2 (Moderate: DeFi will be in the process of being implemented, but won't yet be widespread across the industry), 3 (High: DeFi will be widely used by a majority of financial institutions) 4 (Very high: DeFi will be deeply integrated into nearly all aspects of the financial services industry, with extensive use by most organisations).

Figure 2 illustrates the percentage of experts who perceive DeFi adoption as "High" or "Very high" today versus in 2034. Currently, no experts perceive DeFi adoption as high or very high, reflecting the nascent state of the technology. However, expectations shift dramatically when looking ahead to 2034, with 43.1% of experts anticipating high or very high adoption levels.

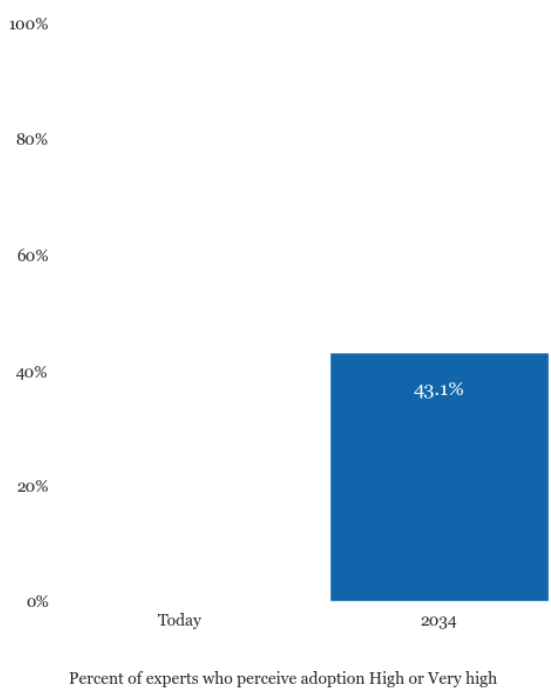

**Figure 2**. Survey evidence on the perceived DeFi adoption today and in 2034. Related survey questions: 1) How would you assess the current levels of DeFi adoption within the financial services industry as a whole? and 2) How would you assess the 2034 levels of DeFi adoption within the financial services industry as a whole? Respondents are asked to rate on a 5-point scale: 0 (Unsure), 1 (Low: DeFi will be used in limited contexts or only by few organizations), 2 (Moderate: DeFi will be in the process of being implemented, but won't yet be widespread across the industry), 3 (High: DeFi will be widely used by a majority of financial institutions) 4 (Very high: DeFi will be deeply integrated into nearly all aspects of the financial services industry, with extensive use by most organisations). The survey is based on 109 expert responses.

This adoption trajectory from 0% to 43.1% mirrors classic emerging market development patterns, where initial scepticism gives way to rapid growth as infrastructure matures and regulatory frameworks stabilize. Table 6 presents the statistical analysis of these adoption perceptions. Panel A shows no significant differences between groups for current adoption (all p > 0.05), with Panel B confirming that all groups report a median adoption level of 1 (Low). This consensus reflects the reality that DeFi, thus far, remains in its early stages within the financial services industry. However, the analysis reveals significant divergence in expectations for 2034. The Kruskal-Wallis tests show significant differences between both expert groups (H = 9.45, p = 0.009) and experience levels (H = 10.63, p = 0.014). Post-hoc analyses reveal that DeFi Industry Practitioners are significantly more optimistic (median = 3) than Course Participants (median = 2), and professionals with over 20 years of experience expect higher adoption (median = 3) than those with 1-5 years of experience (median = 2).

These findings align with my theoretical frameworks. From a NIE perspective, DeFi Industry Practitioners' optimism likely reflects their investment in building new institutional arrangements and their perception of potential efficiency gains from disintermediation. The Dynamic Capabilities perspective helps explain why experienced professionals show greater optimism: Their accumulated experience observing previous financial innovations may enhance their sensing capabilities, allowing them to better recognize DeFi's transformative potential. The contrast between current adoption and future expectations suggests that while DeFi has yet to achieve significant penetration, expert opinion converges on its eventual importance. This indicates that financial services organizations should begin developing relevant capabilities now to prepare for the anticipated transformation. This adoption trajectory mirrors classic emerging market development patterns, where initial scepticism gives way to rapid growth as infrastructure matures and regulatory frameworks stabilize.

**Table 6.** DeFi Adoption Levels

**Panel A.** Statistical Analysis

| Variable | Expert Groups H | Expert Groups p-value | Expert Groups Significant Pairs | Experience Levels H | Experience Levels p-value | Experience Levels Significant Pairs |
|---|---|---|---|---|---|---|
| Adoption today | 0.69 | 0.707 | - | 4.87 | 0.182 | - |
| Adoption 2034 | 9.45 | 0.009** | IP > CP | 10.63 | 0.014* | 20+ > 1-5 |

**Panel B.** Group Medians (IQR)

| Variable | CP | AR | IP | 1-5 yrs | 5-10 yrs | 10-20 yrs | 20+ yrs |
|---|---|---|---|---|---|---|---|
| Adoption today | 1.0 (0.0) | 1.0 (0.0) | 1.0 (0.0) | 1.0 (0.0) | 1.0 (0.0) | 1.0 (0.0) | 1.0 (1.0) |
| Adoption 2034 | 2.0 (1.0) | 2.0 (1.2) | 3.0 (1.0) | 2.0 (0.0) | 2.0 (1.0) | 2.0 (1.0) | 3.0 (1.0) |

Note: ** p < .01, * p < .05. CP = DeFi Course Participants, IP = DeFi Industry Practitioners, AR = DeFi-focused Academic Researchers. IQR = Interquartile Range shown in parentheses. A dash (-) in the Significant Pairs column indicates either no significant differences were found in the overall test or post-hoc pairwise comparisons did not reach significance after Bonferroni correction.

*DeFi platform role*

To elicit expert opinions regarding the role of DeFi platforms I ask, "How do you envision the role of DeFi platforms in shaping the future of the financial services industry by 2034?". Experts are asked to rate on a 6-point scale: 0 (Unsure), 1 (Minimal impact: DeFi platforms will have limited influence, because traditional financial institutions will adapt and innovate to maintain their competitiveness (almost no disruption)), 2 (Niche players: DeFi platforms will occupy niche roles, providing additional services that complement the traditional financial institutions (minor disruption)), 3 (Regulated actors: DeFi platforms will be subject to strict regulations, which will limit their ability to disrupt the banking industry. DeFi platforms will operate independently AND as essential technology and infrastructure services to banks and financial institutions (limited disruption)), 4 (Strategic partners: DeFi platforms and traditional institutions will coexist, forming strategic partnerships (moderate disruption)), 5 (DeFi dominance: DeFi platforms will displace traditional financial services institutions and drive innovation in financial products and services (major disruption)).

Figure 3 depicts experts' views on the potential role of DeFi platforms in the financial ecosystem by 2034. The distribution reveals that most experts expect moderate levels of disruption, with the largest proportion (43%) believing DeFi platforms will serve as niche players complementing traditional finance. The distribution of responses provides additional context: 21% of experts foresee DeFi platforms operating as regulated actors with limited disruptive potential, 16% predict strategic partnerships resulting in moderate disruption, and only 7% expect DeFi platforms to achieve dominance and drive major disruption. Notably, 70% of experts believe DeFi will cause at least limited disruption to the financial services industry by 2034.

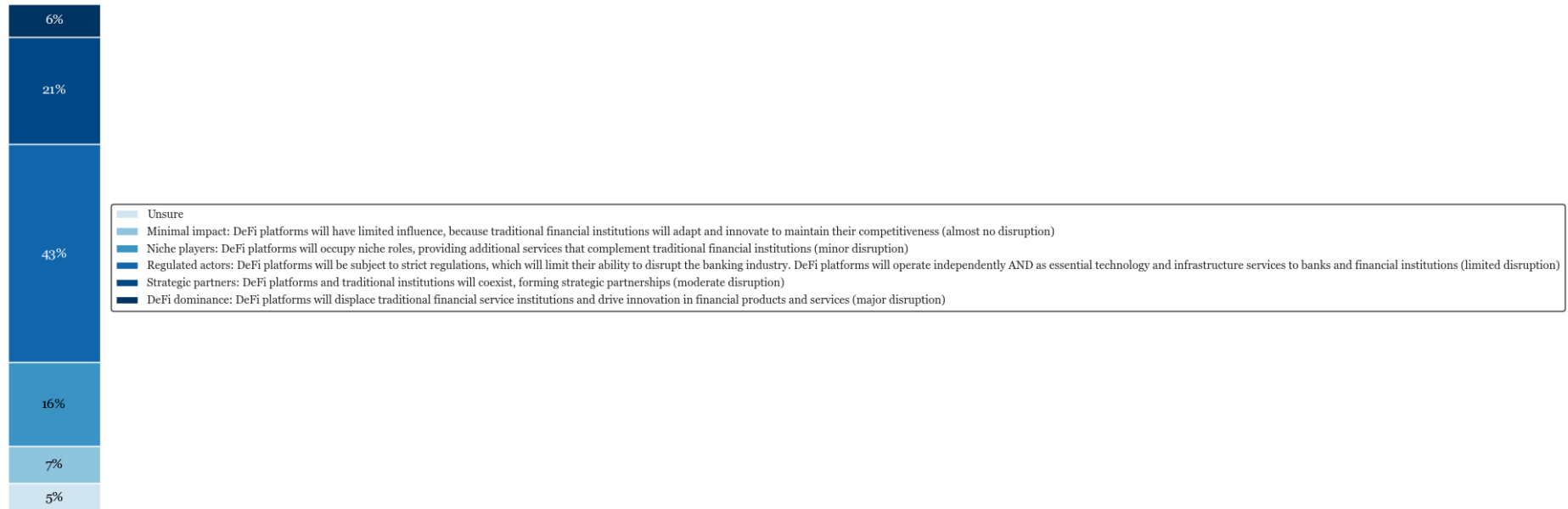

**Figure 3**. Survey evidence on the role of DeFi platforms. Related survey question: How do you envision the role of DeFi platforms in shaping the future of the financial services industry by 2034? Respondents are asked to rate on a 6-point scale: 0 (Unsure), 1 (Minimal impact: DeFi platforms will have limited influence, because traditional financial institutions will adapt and innovate to maintain their competitiveness (almost no disruption)), 2 (Niche players: DeFi platforms will occupy niche roles, providing additional services that complement the traditional financial institutions (minor disruption)), 3 (Regulated actors: DeFi platforms will be subject to strict regulations, which will limit their ability to disrupt the banking industry. DeFi platforms will operate independently AND as essential technology and infrastructure services to banks and financial institutions (limited disruption)), 4 (Strategic partners: DeFi platforms and traditional institutions will coexist, forming strategic partnerships (moderate disruption)), 5 (DeFi dominance: DeFi platforms will displace traditional financial service institutions and drive innovation in financial products and services (major disruption)). The survey is based on 109 expert responses.

Table 7 presents the statistical analysis of these platform role expectations. The results reveal highly significant differences between expert groups (H = 17.60, p < 0.001) and experience levels (H = 12.39, p = 0.006). Post-hoc analyses show that DeFi Industry Practitioners hold significantly more expansive views of DeFi's future role (median = 3) compared to Course Participants (median = 2). Similarly, professionals with 10-20 years and those with over 20 years of experience both expect more substantial platform roles (median = 3) than those with 1-5 years of experience (median = 2).

From a NIE perspective, the expectation that DeFi platforms will predominantly serve as niche players or regulated actors suggests that existing institutional arrangements in financial services possess considerable resilience. The regulatory pressures implied by the "regulated actors" scenario align with institutional isomorphism theory, which predicts that novel arrangements will face pressure to conform to existing regulatory frameworks. The Dynamic Capabilities perspective helps explain the experience-based differences in expectations. Professionals with longer tenure may better recognize how financial innovations typically evolve; that is often starting as disruptive forces but eventually being integrated into existing institutional structures through partnerships or regulatory integration. Their higher expectations for DeFi's role may reflect accumulated understanding about technology adoption cycles in the heavily regulated financial service industry.

These findings suggest that financial services organizations probably should prepare for a future where DeFi platforms play meaningful but likely complementary roles. Rather than wholesale disruption, the expert consensus points toward integration and coexistence, requiring organizations to develop capabilities for partnering with and leveraging DeFi platforms while maintaining their core institutional functions.

The expert consensus on DeFi's platform role is supported by empirical studies of institutional blockchain adoption. Popović et al. (2024) document how established financial institutions achieve operational efficiency through private blockchain implementations, with particularly significant improvements in settlement processes and compliance automation. These findings validate my experts' expectation that DeFi platforms will most likely serve as regulated actors and strategic partners rather than disruptive replacements.

**Table 7.** DeFi Platform Role

**Panel A.** Statistical Analysis

| Variable | Expert Groups H | Expert Groups p-value | Expert Groups Significant Pairs | Experience Levels H | Experience Levels p-value | Experience Levels Significant Pairs |
|---|---|---|---|---|---|---|
| Platform Role | 17.60 | 0.000** | IP > CP | 12.39 | 0.006** | 10-20 > 1-5, 20+ > 1-5 |

**Panel B.** Group Medians (IQR)

| Variable | CP | AR | IP | 1-5 yrs | 5-10 yrs | 10-20 yrs | 20+ yrs |
|---|---|---|---|---|---|---|---|
| Platform Role | 2.0 (1.0) | 3.0 (1.2) | 3.0 (1.0) | 2.0 (2.0) | 3.0 (1.0) | 3.0 (1.0) | 3.0 (1.0) |

Note: ** $p < .01$, * $p < .05$. CP = DeFi Course Participants, IP = DeFi Industry Practitioners, AR = DeFi-focused Academic Researchers. IQR = Interquartile Range shown in parentheses.

*Possible Scenarios*

To elicit detailed opinions about DeFi's future role in financial services, survey participants were presented with Table 8, which outlines five potential scenarios for the financial industry in the context of DeFi's evolution. The first scenario, "Finance as Usual", envisions a continuation of current trends, where financial services firms maintain their traditional competition and innovation progresses only incrementally. The second scenario, "Highly Regulated DeFi" envisions DeFi becoming tightly regulated, offering security and legal certainty to users and financial service providers while limiting its potential for larger-scale disruption. In the third scenario, "DeFi Platform Revolution", DeFi platforms drive a significant transformation in financial services by becoming dominant and providing a wide range of diverse products and services that outpace those of traditional institutions. The fourth scenario, "TradFi Embraces DeFi", imagines traditional financial institutions (TradFi) incorporating DeFi technologies and infrastructure, enabling them to offer improved and accessible services to clients. Finally, the fifth scenario, "Finance for Planet, People, and Common Goods", describes a future where financial institutions and DeFi protocols prioritize human-centred and sustainable practices, focusing on socially responsible activities and investments that create positive social and environmental impacts while generating long-term financial returns.

**Table 8**. Possible scenarios.

| | 1 | 2 | 3 | 4 | 5 |
|---|---|---|---|---|---|
| **Scenarios** | **Finance as usual** | **Highly regulated DeFi** | **DeFi platform revolution** | **TradFi embraces DeFi** | **Finance for planet people and common goods** |
| **Key trends or drivers** | Current trends | Regulations | Independent Platforms | Technology & Innovation | Humans & Society first |
| **Context** | Innovation as usual, current developments continue and financial services firms keep competing amongst themselves | Regulatory requirements mean DeFi has to become highly regulated. This gives security and certainty to users and (de)financial services providers | Innovative and disruptive powers of DeFi platforms lead to a revolution in the financial services industry, with DeFi becoming dominant and providing more diverse products and services only traditional institutions could offer today | With DeFi, current financial services are improved. Institutions see the value in DeFi infrastructure and primitives. Banks support their clients to interact with DeFi components safely and enable access simply. | Financial institutions and DeFi protocols embrace human-centered and sustainable methods separately. They prioritize socially responsible activities and investment opportunities, driving positive impact and boosting long-term returns. |

Then I ask, "How would you rate the following possible scenarios for the year 2034 in relation to DeFi and financial services?". Respondents are asked to rate on a 5-point scale: 0 (Unsure), 1 (Least likely), 2 (Somewhat likely), 3 (Likely), 4 (Most likely). Figure 4 displays the percentage of respondents considering various future DeFi scenarios as likely or very likely. Figure 5 displays the percentage of experts considering various future DeFi scenarios as likely or most likely. The two most anticipated scenarios are "TradFi Embraces DeFi" (66.1%) and "Highly Regulated DeFi" (65.1%), both achieving nearly identical levels of expert endorsement. The "DeFi Platform Revolution" scenario garners moderate support at 51.4%, while "Finance as Usual" (39.4%) and "Finance for Planet, People, and Common Goods" (36.7%) are viewed as the least likely outcomes. The relatively low support for the "Finance as Usual" scenario (39.4%) indicates expert consensus that DeFi will have meaningful impact, while the limited enthusiasm for "DeFi Platform Revolution" (51.4%) suggests scepticism about complete disruption. This pattern reinforces the view that DeFi's future lies in integration and collaboration rather than revolution.

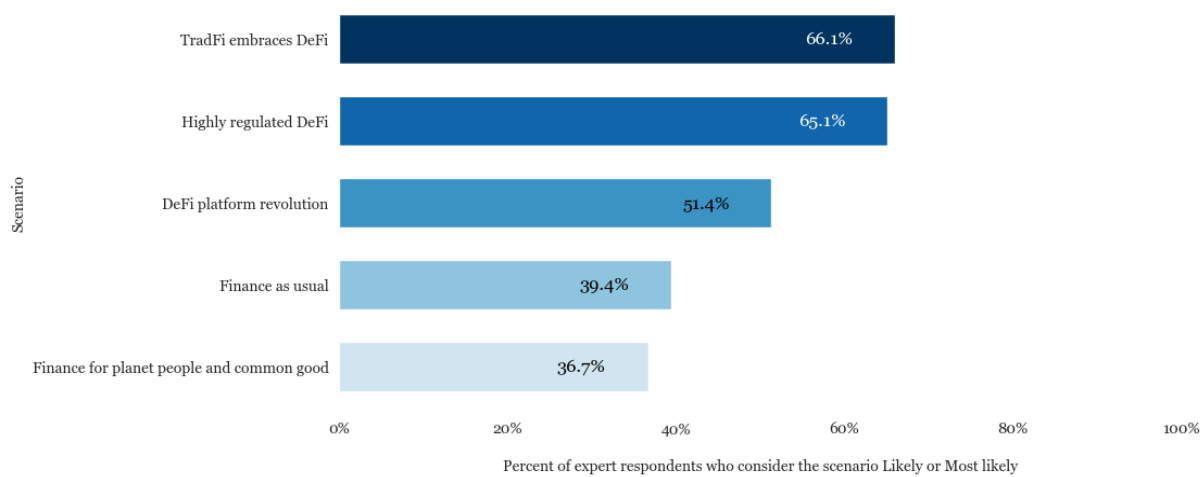

**Figure 4**. Survey evidence on the likelihood of possible scenarios. Related survey question: How would you rate the following possible scenarios for the year 2034 in relation to DeFi and financial services? Respondents are asked to rate on a 5-point scale: 0 (Unsure), 1 (Least likely), 2 (Somewhat likely), 3 (Likely), 4 (Most likely). The survey is based on 109 expert responses.

Table 9 presents the statistical analysis for possible scenarios. Significant differences emerge for two scenarios. For "TradFi Embraces DeFi," both expert groups (H = 10.77, p = 0.005) and experience levels (H = 13.20, p = 0.004) show significant differences. DeFi Industry Practitioners rate this scenario as more likely (median = 3) than Course Participants (median = 3, but with lower overall ratings), and professionals with over 20 years of experience are most optimistic (median = 4) compared to those with 1-5 years (median = 2). The "Highly Regulated DeFi" scenario also shows significant group differences (H = 9.00, p = 0.011), with DeFi Industry Practitioners more convinced of regulatory intervention (median = 3) than Course Participants (median = 2).

The strong expert endorsement of both "TradFi Embraces DeFi" and "Highly Regulated DeFi" scenarios also provides important insights through the theoretical lenses. From a NIE perspective, again these preferred scenarios represent institutional adaptation rather than replacement. The high likelihood assigned to regulatory intervention aligns with institutional theory's prediction that novel financial arrangements face isomorphic pressures to conform to existing regulatory structures. This suggests experts expect DeFi to be integrated into existing institutional frameworks rather than operating outside them completely.

The Dynamic Capabilities perspective casts light on why experienced professionals show greater confidence in the "TradFi Embraces DeFi" scenario. Their accumulated experience may enhance their sensing capabilities, again allowing them to recognize patterns from previous technological integrations in finance. They may better understand how incumbent institutions typically develop dynamic capabilities to adopt and integrate emerging markets and novel technologies into their operating model rather than being displaced by them.

These insights have implications for financial services organizations today. The expert consensus on traditional finance embracing DeFi, combined with expectations of significant regulation, suggests organizations should focus on developing capabilities for DeFi integration within regulatory frameworks. Rather than preparing for disruption or maintaining status quo, the path forward appears to require adaptive strategies that incorporate DeFi technologies while maintaining institutional legitimacy.

**Table 9.** DeFi Scenarios for 2034

**Panel A.** Statistical Analysis

| Variable | Expert Groups H | Expert Groups p-value | Expert Groups Significant Pairs | Experience Levels H | Experience Levels p-value | Experience Levels Significant Pairs |
|---|---|---|---|---|---|---|
| Finance as Usual | 2.81 | 0.246 | - | 6.25 | 0.100 | - |
| Highly Regulated DeFi | 9.00 | 0.011* | IP > CP | 6.22 | 0.101 | - |
| DeFi Platform Revolution | 0.54 | 0.763 | - | 2.24 | 0.525 | - |
| TradFi Embraces DeFi | 10.77 | 0.005** | IP > CP | 13.20 | 0.004** | 20+ > 1-5 |
| Finance for Planet, People & Common Goods | 4.68 | 0.096 | - | 1.98 | 0.576 | - |

**Panel B.** Group Medians (IQR)

| Variable | CP | AR | IP | 1-5 yrs | 5-10 yrs | 10-20 yrs | 20+ yrs |
|---|---|---|---|---|---|---|---|
| Finance as Usual | 2.0 (2.0) | 2.0 (1.2) | 2.0 (2.0) | 3.0 (1.0) | 2.0 (2.0) | 2.0 (2.2) | 1.0 (1.0) |
| Highly Regulated DeFi | 2.0 (1.0) | 3.0 (1.0) | 3.0 (1.0) | 3.0 (1.0) | 3.0 (1.8) | 3.0 (1.2) | 3.0 (2.0) |
| DeFi Platform Revolution | 3.0 (1.0) | 2.0 (1.0) | 3.0 (1.0) | 2.0 (1.0) | 2.0 (1.8) | 3.0 (1.2) | 3.0 (1.0) |
| TradFi Embraces DeFi | 3.0 (1.0) | 3.5 (1.2) | 3.0 (1.0) | 2.0 (1.0) | 3.0 (2.0) | 3.0 (2.0) | 4.0 (1.0) |
| Finance for Planet, People & Common Goods | 2.0 (2.0) | 1.0 (1.2) | 2.0 (2.0) | 2.0 (2.0) | 2.0 (1.0) | 2.0 (2.0) | 3.0 (2.0) |

Note: ** p < .01, * p < .05. CP = DeFi Course Participants, IP = DeFi Industry Practitioners, AR = DeFi-focused Academic Researchers. IQR = Interquartile Range shown in parentheses. A dash (-) in the Significant Pairs column indicates either no significant differences were found in the overall test or post-hoc pairwise comparisons did not reach significance after Bonferroni correction.

## 6.5    Impact on business areas

Next, I am interested in the impact of DeFi on different business areas and processes of financial services firms. I ask: "How do you envision the impact and role of DeFi in these business areas or processes within the banking and financial services industry by the year 2034?". Experts are asked to rate on a 5-point scale: 0 (Unsure), 1 (Minimal), 2 (Low), 3 (Moderate), 4 (High).

Figure 5 shows the percentage of experts who believe DeFi will have a moderate or high impact on various business areas. Risk management emerges as the most impacted area, with 86.2% of experts anticipating significant changes. Data analysis follows closely at 82.6%, with Operations (79.8%), Research and Development (78.9%), and Finance and Accounting (77.1%) also expected to experience substantial impact. Management and decision-making processes show moderate expected impact at 67.0%, while fewer experts anticipate significant changes in Marketing and Sales (49.5%), Other business areas (42.2%), and HR and Recruitment (37.6%).

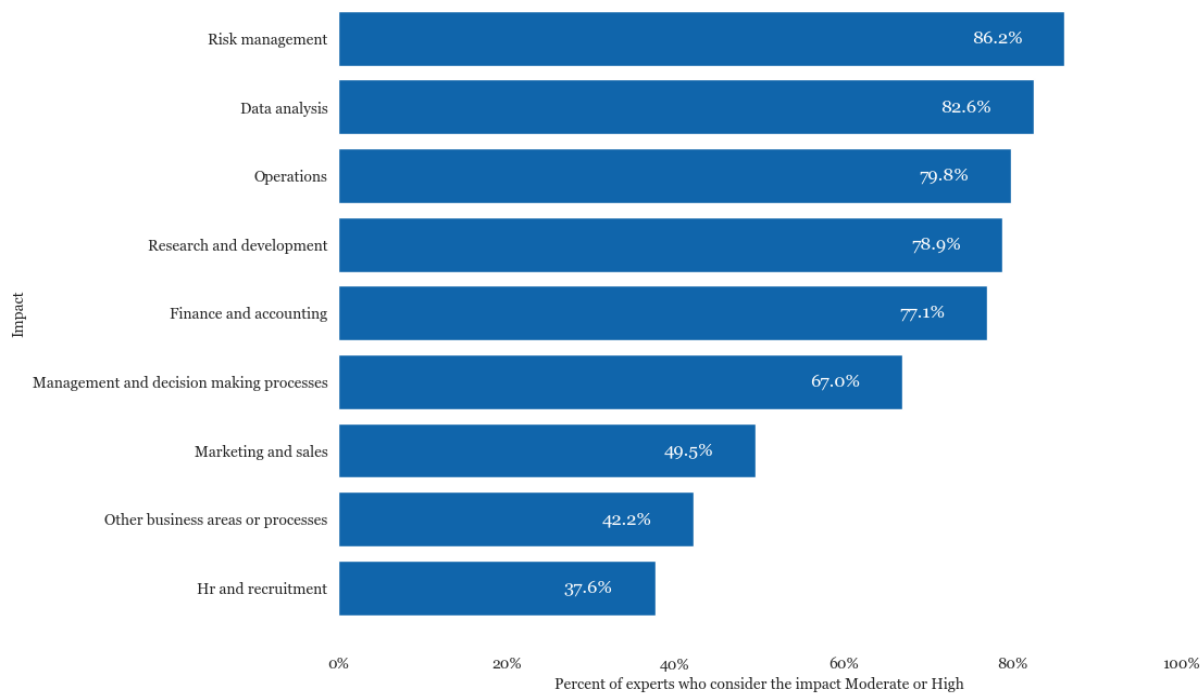

**Figure 5**. Survey evidence on impact of DeFi on business areas in financial services. Related survey question: How do you envision the impact and role of DeFi in these business areas or processes within the banking and financial services industry by the year 2034? Respondents are asked to rate on a 5-point scale: 0 (Unsure), 1 (Minimal), 2 (Low), 3 (Moderate), 4 (High). The survey is based on 109 financial services expert responses.

Table 10 presents the statistical analysis of DeFi's expected impact across business areas. Significant differences between expert groups emerge for Risk Management (H = 8.07, p = 0.018) and Finance and Accounting (H = 8.01, p = 0.018), with DeFi Industry Practitioners consistently expecting higher impact (median = 4) than Course Participants (median = 3) in both areas. Experience levels reveal even more pronounced differences: Operations shows highly significant variation (H = 14.69, p = 0.002), with professionals having 20+ years of experience expecting much higher impact (median = 4) than those with 1-5 years or 5-10 years (median = 3). Similar patterns appear for HR and Recruitment (H = 13.48, p = 0.004), Risk Management (H = 8.86, p = 0.031), and Management and Decision Making (H = 9.33, p = 0.025).

From a NIE perspective, the high expected impact on Risk Management reflects the fundamental challenge DeFi poses to existing risk governance mechanisms in current organizations. Traditional financial institutions have developed sophisticated risk management frameworks over decades, but DeFi introduces novel and idiosyncratic risks, for example smart contract vulnerabilities and scaling-related issues that existing institutional arrangements struggle to address. The significant group differences suggest that DeFi Industry Practitioners, who directly encounter these challenges, appreciate the institutional adaptations required more.

The Dynamic Capabilities lens helps explain the pronounced experience-based differences, particularly for Operations and HR. Senior professionals with 20+ years of experience may better recognize how industry shifts require comprehensive organizational transformation. Their higher impact expectations across multiple business areas suggest they understand that adopting DeFi isn't merely a technical challenge but requires reconfiguring operational processes, developing new human capital competencies, and transforming decision-making structures. The relatively lower expected impact on Marketing and Sales and HR and Recruitment might reflect a view that DeFi primarily transforms back-office and risk functions rather than customer-facing activities. However, the significant experience-based differences for HR suggest that senior professionals recognize the human capital challenges ahead.

These findings indicate that financial services organizations should prepare for comprehensive transformation across core business functions, with particular attention to risk management and operational processes. The expertise gap between current and required capabilities appears substantial, suggesting that organizations must begin developing relevant competencies across multiple business areas to effectively interact with and integrate DeFi.

The high expected impact on risk management also aligns with emerging frameworks in the literature. Adamyk et al. (2025) document novel risk management tools for DeFi, while Aquilina et al. (2024) identify unique risks including composability risk and oracle dependencies that traditional frameworks are yet to address comprehensively. The Basel Committee's prudential treatment requires 1250% risk weighting for unbacked crypto assets, explaining why institutions focus on tokenized traditional assets (BCBS, 2022) thus far. This regulatory reality supports my finding that risk management transformation is the most critical business area impact.

**Table 10.** DeFi Impact on Business Areas

**Panel A.** Statistical Analysis

| Variable | Expert Groups H | Expert Groups p-value | Expert Groups Significant Pairs | Experience Levels H | Experience Levels p-value | Experience Levels Significant Pairs |
|---|---|---|---|---|---|---|
| Data Analysis | 0.07 | 0.968 | - | 3.86 | 0.277 | - |
| Finance and Accounting | 8.01 | 0.018* | IP > CP | 6.58 | 0.086 | - |
| HR and Recruitment | 2.39 | 0.303 | - | 13.48 | 0.004** | 20+ > 5-10 |
| Marketing and Sales | 0.34 | 0.845 | - | 9.30 | 0.026* | - |
| Operations | 1.38 | 0.502 | - | 14.69 | 0.002** | 20+ > 1-5, 20+ > 5-10 |
| Research and Development | 2.57 | 0.277 | - | 4.14 | 0.247 | - |
| Risk Management | 8.07 | 0.018* | IP > CP | 8.86 | 0.031* | 20+ > 1-5 |
| Management and Decision Making | 1.27 | 0.530 | - | 9.33 | 0.025* | 10-20 > 1-5 |
| Other Business Areas | 5.87 | 0.053 | - | 6.83 | 0.077 | - |

**Panel B.** Group Medians (IQR)

| Variable | CP | AR | IP | 1-5 yrs | 5-10 yrs | 10-20 yrs | 20+ yrs |
|---|---|---|---|---|---|---|---|
| Data Analysis | 4.0 (1.0) | 4.0 (1.0) | 3.0 (1.0) | 3.0 (1.0) | 3.0 (2.0) | 4.0 (1.0) | 4.0 (1.0) |
| Finance and Accounting | 3.0 (2.0) | 3.0 (2.0) | 4.0 (1.0) | 3.0 (1.0) | 3.5 (2.0) | 3.5 (1.0) | 3.0 (1.0) |
| HR and Recruitment | 2.0 (1.0) | 2.0 (0.5) | 2.0 (2.0) | 2.0 (2.0) | 2.0 (1.0) | 2.0 (1.0) | 3.0 (1.0) |
| Marketing and Sales | 3.0 (1.0) | 2.0 (1.0) | 2.0 (1.0) | 2.0 (2.0) | 2.0 (2.0) | 3.0 (1.0) | 3.0 (1.0) |
| Operations | 3.5 (1.8) | 3.0 (1.2) | 3.0 (1.0) | 3.0 (2.0) | 3.0 (1.8) | 4.0 (1.0) | 4.0 (1.0) |
| Research and Development | 3.0 (2.0) | 3.0 (1.0) | 3.0 (1.0) | 3.0 (2.0) | 3.0 (2.0) | 3.0 (1.0) | 3.0 (1.0) |
| Risk Management | 3.0 (1.0) | 3.5 (1.0) | 4.0 (1.0) | 3.0 (3.0) | 4.0 (1.0) | 4.0 (1.0) | 4.0 (1.0) |
| Management and Decision Making | 3.0 (2.0) | 3.0 (1.5) | 3.0 (1.0) | 2.0 (1.0) | 3.0 (1.0) | 3.0 (1.0) | 3.0 (0.8) |
| Other Business Areas | 2.5 (2.8) | 0.0 (2.2) | 2.0 (3.0) | 2.0 (3.0) | 0.0 (2.0) | 2.0 (3.0) | 3.0 (2.8) |

Note: ** p < .01, * p < .05. CP = DeFi Course Participants, IP = DeFi Industry Practitioners, AR = DeFi-focused Academic Researchers. IQR = Interquartile Range shown in parentheses. A dash (-) in the Significant Pairs column indicates either no significant differences were found in the overall test or post-hoc pairwise comparisons did not reach significance after Bonferroni correction.

*6.6      DeFi issues and problems*

Here, I investigate the importance of DeFi issues and problems by asking: "How important do you think these DeFi-related issues and potential problems are for the financial services industry until 2034?". Experts are asked to rate on a 5-point scale: 0 (Unsure), 1 (Not important), 2 (Moderately important), 3 (Important), 4 (Very important).

Figure 6 highlights the issues that experts consider important or very important for DeFi adoption. Security tops the list at 90.8%, followed closely by Data Management and Privacy (87.2%) and Regulatory Challenges (81.7%). Other significant concerns include Tech Infrastructure and Legacy System Upgradability (81.7%), Unknown and Known High Risks (73.4%), and Transparency (73.4%). Additionally, 67.9% cite Lack of DeFi Expertise and 61.5% identify Lack of DeFi Strategy as important barriers. Lower-ranked but notable concerns include Lack of Funding/Budget (55.0%), Job Losses or Upskilling (40.4%), Impact on Environment (37.6%), and Loss of Human Touch in Services (36.7%). These concerns reflect typical emerging market risks: security and regulatory challenges parallel political risk in geographic emerging markets, while technical infrastructure gaps mirror physical infrastructure limitations in developing economies.

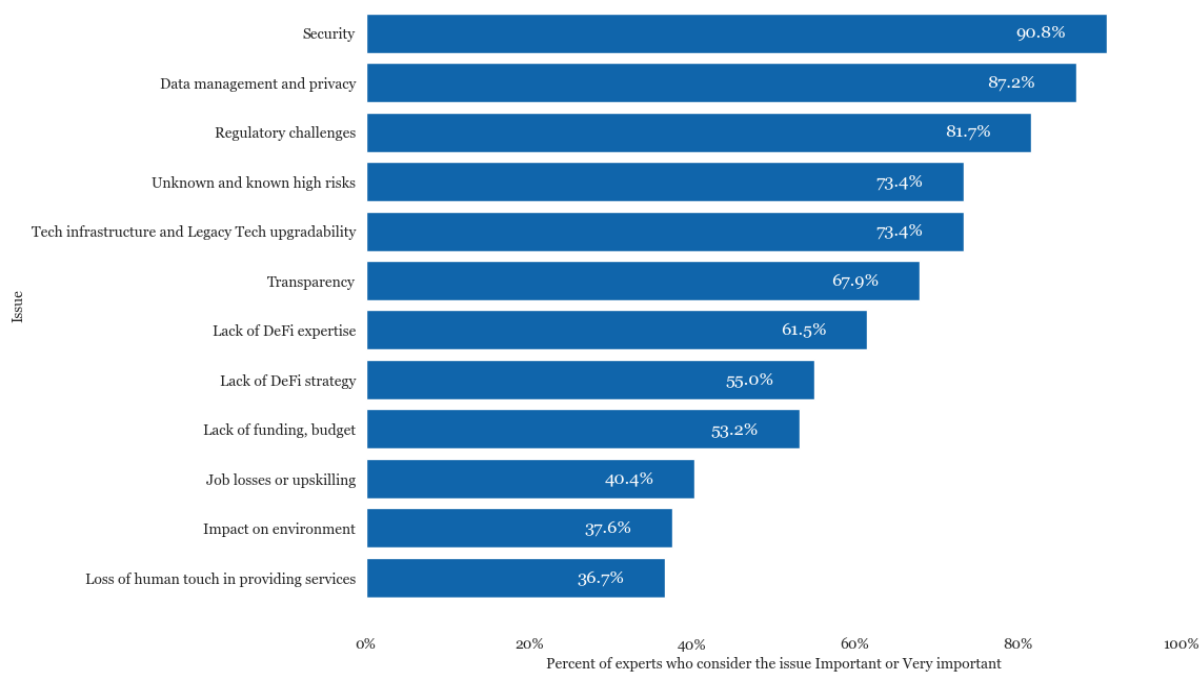

**Figure 6**. Survey evidence on DeFi-related issues and problems. Related survey question: How important do you think these DeFi-related issues and potential problems are for the financial services industry until 2034? Respondents are asked to rate on a 5-point scale: 0 (Unsure), 1 (Not important), 2 (Moderately important), 3 (Important), 4 (Very important). The survey is based on 109 financial services expert responses.

These concerns reflect typical emerging market risks: security and regulatory challenges parallel political risk in geographic emerging markets, while technical infrastructure gaps mirror physical infrastructure limitations in developing economies.

Table 11 reveals striking differences in how expert groups perceive these challenges. The largest divergence appears for Loss of Human Touch (H = 11.58, p = 0.003) and Job Losses or Upskilling (H = 18.62, p < 0.001), where Course Participants rate these concerns significantly higher (median = 3) than both Academic Researchers (median = 1) and DeFi Industry Practitioners (median = 2). Environmental Impact shows similar patterns (H = 13.45, p = 0.001), with Course Participants more concerned (median = 3) than other groups (median = 2). Conversely, for Regulatory Challenges, DeFi Industry Practitioners express greater concern (median = 4) than Course Participants (median = 3, H = 6.36, p = 0.042). Lack of Funding/Budget reveals an interesting pattern where both Course Participants and DeFi Industry Practitioners (median = 3) rate it as more important than Academic Researchers (median = 1, H = 17.02, p < 0.001).

From a NIE perspective, the universal concern about security and regulatory challenges reflects the fundamental tension between DeFi's trustless systems and institutional trust requirements. Traditional financial institutions exist partly to solve trust problems, but DeFi's trustless architecture relying on cryptographic proofs allows any actor to participate without institutional vetting. This creates what NIE would identify as a governance vacuum that heightens security risks and regulatory uncertainty. The higher concern among DeFi Industry Practitioners regarding regulation may suggest they worry more about the institutional barriers to DeFi adoption.

The Dynamic Capabilities perspective highlights the divergent views on human-centred concerns. Course Participants' significantly higher worry about job losses and loss of human touch may reflect anxiety about capability obsolescence. In contrast, DeFi Industry Practitioners and Academic Researchers may view these changes as capability evolution rather than replacement. The universal recognition of expertise and strategy gaps as important barriers aligns with DCT where organizations must develop new sensing, seizing, and transforming capabilities specific to the DeFi phenomenon.

These insights have practical implications, too. The interplay between security, privacy, and regulatory compliance presents a complex challenge. Organizations must balance anti-money laundering requirements with privacy protection laws like GDPR and PDPA. The high importance placed on expertise and strategy gaps indicates an urgent need for comprehensive DeFi education and strategic planning. Interestingly, the relatively lower concern about environmental impact may reflect overall awareness of energy-efficient consensus mechanisms like Proof-of-Stake. However, the significant group differences on human-centred issues suggest that change management and workforce development strategies must address varying stakeholder concerns about DeFi's impact on employment and service delivery models.

**Table 11.** DeFi-related Issues and Problems

**Panel A.** Statistical Analysis

| Variable | Expert Groups H | Expert Groups p-value | Expert Groups Significant Pairs | Experience Levels H | Experience Levels p-value | Experience Levels Significant Pairs |
|---|---|---|---|---|---|---|
| Ethical Issues | 3.88 | 0.143 | - | 8.08 | 0.044* | - |
| Data Management & Privacy | 0.32 | 0.852 | - | 8.61 | 0.035* | - |
| Transparency | 1.37 | 0.504 | - | 1.91 | 0.590 | - |
| Lack of DeFi Expertise | 0.66 | 0.718 | - | 0.73 | 0.867 | - |
| Security | 5.87 | 0.053 | - | 4.92 | 0.178 | - |
| Regulatory Challenges | 6.36 | 0.042* | IP > CP | 2.81 | 0.421 | - |
| Unknown & Known High Risks | 1.66 | 0.437 | - | 8.40 | 0.038* | - |
| Loss of Human Touch | 11.58 | 0.003** | CP > AR, CP > IP | 7.88 | 0.049* | - |
| Job Losses or Upskilling | 18.62 | 0.000** | CP > AR, CP > IP | 10.06 | 0.018* | 1-5 > 5-10, 20+ > 5-10 |
| Lack of DeFi Strategy | 2.59 | 0.274 | - | 3.46 | 0.326 | - |
| Tech Infrastructure & Legacy Tech | 1.85 | 0.396 | - | 5.07 | 0.167 | - |
| Impact on Environment | 13.45 | 0.001** | CP > AR, CP > IP | 1.32 | 0.723 | - |
| Lack of Funding/Budget | 17.02 | 0.000** | CP > AR, IP > AR | 2.76 | 0.431 | - |

**Panel B.** Group Medians (IQR)

| Variable | CP | AR | IP | 1-5 yrs | 5-10 yrs | 10-20 yrs | 20+ yrs |
|---|---|---|---|---|---|---|---|
| Ethical Issues | 3.0 (2.0) | 2.5 (1.0) | 2.0 (1.0) | 3.0 (1.0) | 2.0 (1.8) | 2.5 (2.0) | 3.0 (2.0) |
| Data Management & Privacy | 4.0 (1.0) | 4.0 (1.2) | 4.0 (1.0) | 3.0 (1.0) | 3.0 (1.8) | 4.0 (1.0) | 4.0 (0.8) |
| Transparency | 3.0 (1.0) | 3.0 (2.0) | 3.0 (2.0) | 3.0 (0.0) | 3.0 (2.0) | 3.0 (2.0) | 3.0 (1.0) |
| Lack of DeFi Expertise | 3.0 (1.0) | 3.0 (2.0) | 3.0 (2.0) | 3.0 (1.0) | 3.0 (2.0) | 3.0 (2.0) | 3.0 (2.0) |
| Security | 4.0 (1.0) | 3.5 (1.0) | 4.0 (0.0) | 4.0 (1.0) | 4.0 (1.0) | 4.0 (1.0) | 4.0 (0.0) |
| Regulatory Challenges | 3.0 (1.0) | 4.0 (1.0) | 4.0 (1.0) | 3.0 (1.0) | 3.5 (2.0) | 4.0 (1.0) | 4.0 (1.0) |
| Unknown & Known High Risks | 3.0 (1.8) | 3.0 (1.2) | 3.0 (1.0) | 3.0 (1.0) | 3.0 (1.8) | 3.0 (2.0) | 3.0 (1.0) |
| Loss of Human Touch | 3.0 (1.0) | 1.0 (1.0) | 2.0 (1.0) | 3.0 (2.0) | 1.0 (1.0) | 2.0 (2.0) | 2.0 (1.0) |
| Job Losses or Upskilling | 3.0 (2.0) | 1.0 (1.2) | 2.0 (1.0) | 2.0 (2.0) | 1.0 (1.0) | 2.0 (2.0) | 2.5 (1.0) |
| Lack of DeFi Strategy | 3.0 (2.0) | 3.0 (2.0) | 3.0 (1.0) | 3.0 (1.0) | 3.0 (1.0) | 3.0 (1.2) | 3.0 (1.8) |
| Tech Infrastructure & Legacy Tech | 3.0 (1.0) | 3.0 (2.0) | 3.0 (1.0) | 3.0 (1.0) | 3.0 (0.0) | 3.0 (2.0) | 3.0 (1.0) |
| Impact on Environment | 3.0 (1.8) | 2.0 (1.0) | 2.0 (1.0) | 2.0 (2.0) | 2.0 (1.8) | 2.0 (2.0) | 2.0 (1.0) |
| Lack of Funding/Budget | 3.0 (1.0) | 1.0 (1.0) | 3.0 (1.0) | 2.0 (1.0) | 3.0 (1.0) | 3.0 (1.2) | 3.0 (2.0) |

Note: ** p < .01, * p < .05. CP = DeFi Course Participants, IP = DeFi Industry Practitioners, AR = DeFi-focused Academic Researchers. IQR = Interquartile Range shown in parentheses. A dash (-) in the Significant Pairs column indicates either no significant differences were found in the overall test or post-hoc pairwise comparisons did not reach significance after Bonferroni correction.

## 6.7    Customer service

The next question relates to customer service: "How do you envision the impact and role of DeFi in customer service and experience within the financial services industry by the year 2034?". Experts are asked to rate on a 5-point scale: 0 (Unsure), 1 (Minimal: DeFi will have almost no influence on customer service), 2 (Low: DeFi will make minor improvements but not fundamentally altering the customer experience), 3 (Moderate: DeFi will play a significant role in improving customer service in the banking and financial services industry, streamlining processes and delivering better support), 4 (High: DeFi will revolutionise customer service by automating almost all or most interactions and by providing personalised, efficient client experiences).

Figure 7 illustrates experts' views on the potential impact of DeFi on customer service in financial services. 37%, believe that DeFi will play a "moderate" role in banking and financial services' customer service by streamlining processes and enhancing support. Another 33% foresee a "low" impact, with DeFi contributing only minor improvements without fundamentally altering the customer experience. Meanwhile, 16% of experts expect DeFi to have a "high" impact, revolutionizing customer service by automating interactions and delivering personalized, efficient experiences. A smaller percentage, 7%, anticipate "minimal" influence, while 4% remain "unsure" about DeFi's effect on customer service. In summary, over half of the experts interpret a rise of DeFi as positive or an opportunity to improve customer service.

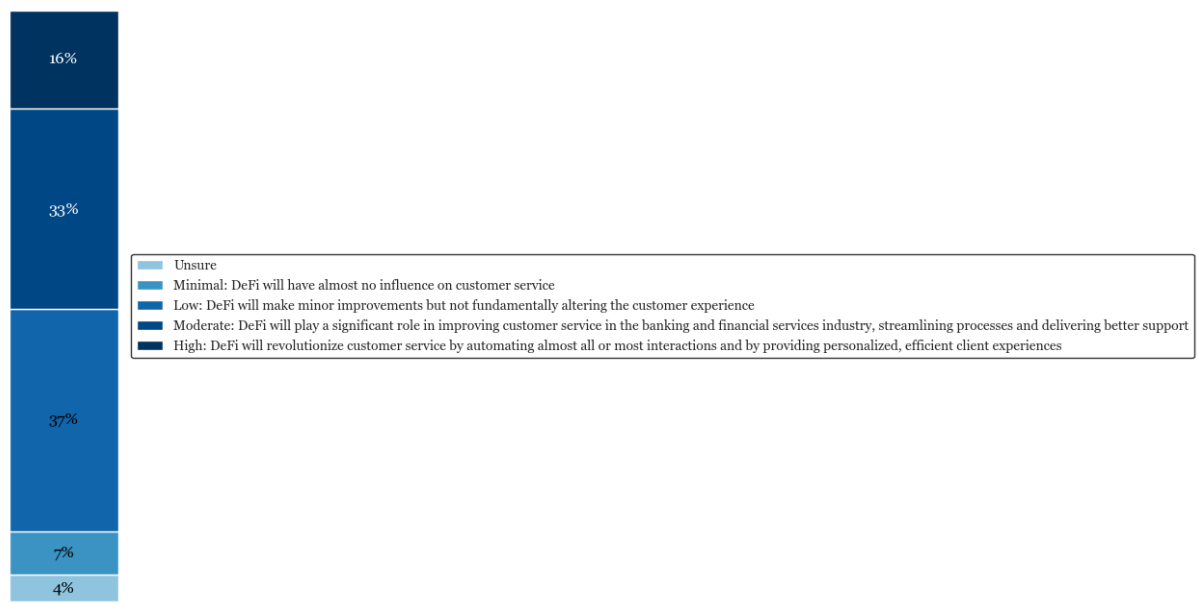

**Figure 7**. DeFi and Customer service. Related survey question: How do you envision the impact and role of DeFi in customer service and experience within the financial services industry by the year 2034? Respondents are asked to rate on a 5-point scale: 0 (Unsure), 1 (Minimal: DeFi will have almost no influence on customer service), 2 (Low: DeFi will make minor improvements but not fundamentally altering the customer experience ), 3 (Moderate: DeFi will play a significant role in improving customer service in the banking and financial services industry, streamlining processes and delivering better support), 4 (High: DeFi will revolutionise customer service by automating almost all or most interactions and by providing personalised, efficient client experiences). The survey is based on 109 financial services expert responses.

Table 12 presents the statistical analysis of expected customer service impact. While no significant differences emerge between expert groups (H = 3.51, p = 0.173), experience levels show significant variation (H = 9.17, p = 0.027). Professionals with over 20 years of experience anticipate higher impact on customer service (median = 3) compared to those with 1-5 years of experience (median = 2). Notably, the most experienced professionals show complete consensus in their assessment, with an IQR of 0.0.

From a NIE perspective, the moderate expected impact on customer service reflects the enduring importance of trust-based relationships in financial services. While DeFi can streamline processes through automation and smart contracts, the institutional role of financial service providers in managing complex customer needs and providing trusted advice may be difficult to fully disintermediate. The lack of significant differences between expert groups suggests broad agreement that customer-facing activities will undergo evolution rather than revolution.

The Dynamic Capabilities lens helps explain why experienced professionals anticipate greater customer service transformation. Their longer tenure may provide superior sensing capabilities to recognize how industry innovations eventually reshape customer interactions, even in relationship-driven industries. They may recall how online banking, mobile apps, and robo-advisors that were initially seen as minor enhancements, fundamentally changed customer service expectations and delivery models. Their higher expectations suggest DeFi could follow a similar trajectory.

Insights here indicate that while DeFi is expected to enhance customer service through process improvements and automation, it is unlikely to eliminate the human element entirely. Financial services organizations could focus on developing capabilities that blend DeFi's benefits with maintained personal relationships. DeFi appears poised to augment rather than replace customer service functions, requiring organizations to thoughtfully integrate DeFi tools and platforms while preserving the trusted advisor role that many customers still value, even if delivered online.

**Table 12.** DeFi and Customer Service

**Panel A.** Statistical Analysis

| Variable | Expert Groups H | Expert Groups p-value | Expert Groups Significant Pairs | Experience Levels H | Experience Levels p-value | Experience Levels Significant Pairs |
|---|---|---|---|---|---|---|
| Customer Service Impact | 3.51 | 0.173 | - | 9.17 | 0.027* | 20+ > 1-5 |

**Panel B.** Group Medians (IQR)

| Variable | CP | AR | IP | 1-5 yrs | 5-10 yrs | 10-20 yrs | 20+ yrs |
|---|---|---|---|---|---|---|---|
| Customer Service Impact | 2.0 (1.0) | 2.0 (1.2) | 3.0 (1.0) | 2.0 (1.0) | 2.0 (1.0) | 3.0 (1.0) | 3.0 (0.0) |

Note: ** p < .01, * p < .05. CP = DeFi Course Participants, IP = DeFi Industry Practitioners, AR = DeFi-focused Academic Researchers. IQR = Interquartile Range shown in parentheses.

## 6.8 Competencies

After discussing the impact of DeFi on financial services, I now turn to potentially required competencies, or what we can call DeFi literacy. Before asking the competency-related question, the online survey shows participants Table 13. This table presents a structured framework to understand the competencies required by organizations to engage with DeFi. There are three levels of competencies. Level 1 outlines the general capabilities an organization must develop; in my case, that is just DeFi as a header. Level 2 describes the organizational competency categories at an organizational level. It introduces six broad categories of organizational capabilities. The categories are sector-specific domain expertise, technological competencies, cognitive competencies, interactional competencies, strategic and organizational competencies, and ethical and societal competencies. Level 3 provides detailed examples of the specific skills and knowledge required under each category. Under sector-specific competencies, the framework suggests that organizations must develop knowledge of DeFi infrastructure, including blockchain, smart contract platforms, oracles, stablecoins, and the concept of decentralized applications (DApps). Additionally, they must understand DeFi primitives, including transactions, token types, burn/mint mechanisms, bonding curves, incentives, staking/slashing mechanisms, fees, and swaps. Further, organizations may need to be familiar with DeFi applications such as automated market makers, borrowing/lending platforms, derivatives, insurance, and tokenization. They must also be able to identify opportunities such as financial inclusion, composability, efficiency, and centralized control, as well as risks such as impermanent loss, smart contract risk, including MEV or miner/maximal extractable value risk as described in Daian et al. (2019), oracle risk, governance risk, and scaling challenges. Organizations need technological competencies, including hardware, software, and network-related proficiency. They should also understand how DeFi software is developed, including being familiar with relevant software development kits (SDKs) and Solidity (the primary programming language for smart contracts on Ethereum). Organizations must also be able to manage data and security, ensuring privacy-preserving data practices.

The cognitive competencies required to interact with the DeFi phenomenon, include fostering organizational learning and develop skills in emotion recognition, problem-solving, and decision-making. These competencies appear to be critical for navigating the rapid and often complex developments in DeFi. Additionally, financial services firms should have research and development capabilities to innovate and remain competitive. Interactional competencies emphasize the importance of managing relationships with customers and other companies, protocols, and also decentralized autonomous organizations (DAOs) as described in Liebau and Oh (2024). Organizations may also need to continuously upskill their workforce to adapt to new technologies and changes within the DeFi ecosystem. They should also focus on aligning employee opportunities with DeFi developments and fostering empathy in interactions. For strategic and organizational competencies, organizations may have to demonstrate strong decision-making, leadership and be able to identify opportunities related to financial inclusion, efficiency, and interoperability within DeFi. They may need strategic vision, be able to manage DeFi-related portfolios of work well, implement processes, and effectively manage expectations regarding DeFi. Innovation management appears critical, as may be the ability to participate in governance structures like DAOs. Additionally, organizations may benefit from the ability to co-pilot DeFi concepts, such as using automated market makers (AMMs) for decentralized exchanges (e.g., AMMs for EMFX, a decentralized trading platform). Lastly, ethical and societal competencies may require organizations to understand the broader impact of DeFi on societal structures and manage civic, ethical, and legal responsibilities. Other focus areas include regulatory compliance and the promotion of sustainable development. Importantly, organizations must also safeguard data privacy and ensure confidentiality in all interactions and processes.

**Table 13**. DeFi competency overview.

| Level 1 *(General capabilities of an organization)* | Level 2 *(Key categories of DeFi competencies on an organizational level)* | Level 3 *(Specific examples of key DeFi competencies)* |
|---|---|---|
| DeFi competencies | Sector-specific domain competencies | - Organization or industry-specific types of knowledge and experience:<br>  - DeFi infrastructure (Blockchain, Smart Contract (Platforms), Oracles, Stablecoins, DApps, etc.)<br>  - DeFi primitives (Transactions, Token types, Burn/Mint, Bonding curves, Incentives, Staking/Slashing, Fees, Swaps, etc.)<br>  - DeFi Apps (Automated Market Makers, DeFi Borrowing/Lending, Derivatives, Insurance, Tokenization, etc.)<br>  - Identification of opportunities (Financial inclusion, Composability, Efficiency, Centralized Control, etc.)<br>  - Identification of risks (MEV, Impermanent Loss, Smart Contract risk, Oracle risk, Governance risk, Scaling risk, etc.) |
| | Technological competencies | - Technologies and other essential resources<br>  - Technology infrastructure (hardware, software, networks)<br>  - Development of DeFi software (SDKs, Solidity, etc.)<br>  - Data and data management (including privacy-preserving methods)<br>  - Security management" |
| | Cognitive competencies | - Organizational learning<br>  - Emotion recognition<br>  - Problem-solving, decision making<br>  - Research and Development capability" |
| | Interactional competencies | - Managing relationships<br>  - Interaction with customers, companies, protocols, DAOs<br>  - Continuous upskilling of the current workforce<br>  - Aligning employee opportunities with DeFi systems<br>  - Empathy |
| | Strategic and organizational competencies | - Managerial decision making and leadership<br>  - Identification of opportunities (Financial Inclusion, efficiency, interoperability)<br>  - Strategic vision and decisions<br>  - Coordination of (portfolios of) work<br>  - Implementation and management of processes<br>  - Managing and achieving expectations with regards to DeFi<br>  - Design capability and strategy<br>  - Innovation Management<br>  - Governance (e.g. ability to participate in a DAO)<br>  - Co-piloting DeFi concepts (e.g. using an AMM for EMFX) |
| | Ethical and societal competencies | - Impact of DeFi on societal structures and rules<br>  - Managing civic, ethical, and legal responsibilities<br>  - Regulatory compliance<br>  - Sustainable development<br>  - Data sharing and privacy/confidentiality |

I ask participants: "How would you evaluate the importance of these DeFi competencies for financial services organisations today and in 2034?". Experts are asked to rate on a 5-point scale: 0 (Unsure), 1 (Not important), 2 (Moderately important), 3 (Important), 4 (Very important).

Figure 8 illustrates experts' perceptions of the importance of various DeFi competencies today and in 2034. Experts expect competencies across all categories to increase significantly by 2034. Strategic competencies are seen as increasingly important, rising from 59.6% today to 84.4% in 2034. Similarly, Sector-specific domain expertise grows from 65.1% to 81.7%, and Technological competencies rise from 62.4% to 78.9%. Other categories, such as Ethical and societal competencies and Interactional competencies, also see notable increases, from 58.7% to 78.0% and 45.0% to 77.1%, respectively. Cognitive competencies increase in importance from 48.6% to 75.2%. The "Other" category sees a more modest rise from 17.4% to 38.5%, indicating that emerging or miscellaneous competencies will play a growing, but still relatively minor, role in the DeFi landscape. The increase in required competencies represents typical emerging market entry barriers, where organizations must develop entirely new capabilities to operate effectively in unfamiliar institutional environments.

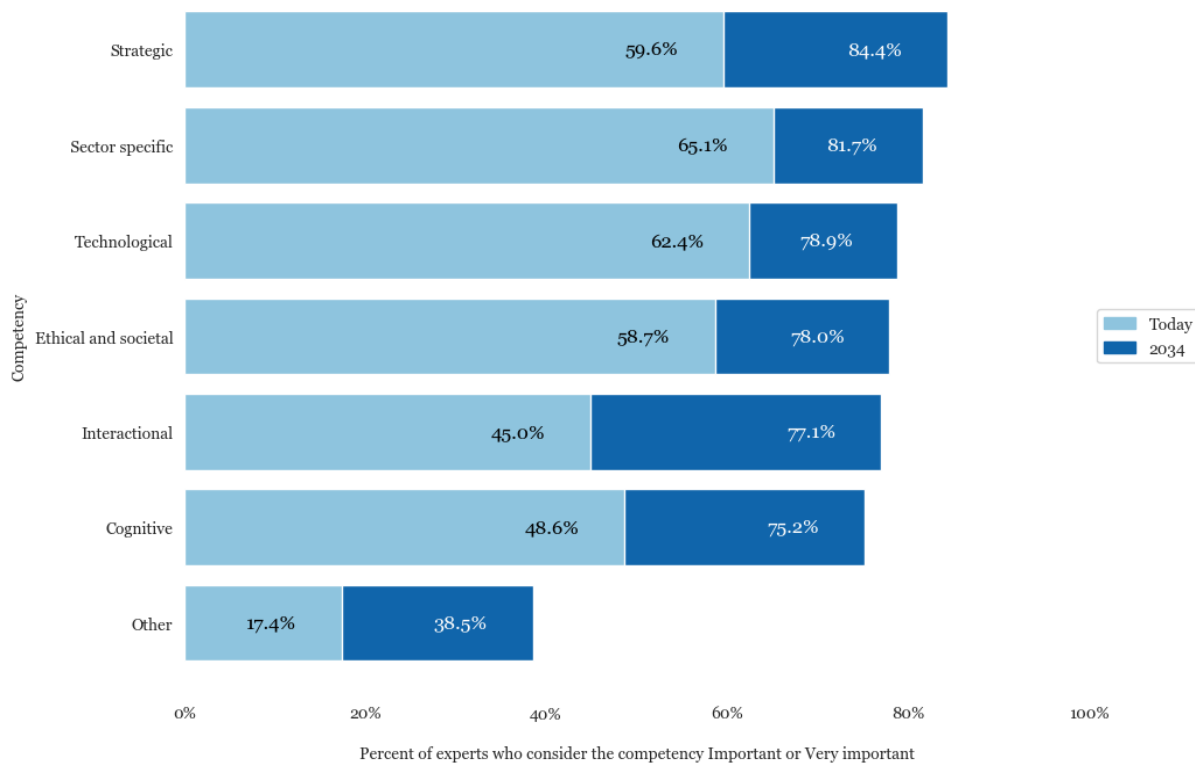

**Figure 8**. Survey evidence on the perceived importance of DeFi competencies today and in 2034. Related survey question: How would you evaluate the importance of these DeFi competencies for financial services organisations today and in 2034? Respondents are asked to rate on a 5-point scale: 0 (Unsure), 1 (Not important), 2 (Moderately important), 3 (Important), 4 (Very important). The survey is based on 109 financial services expert responses.

The increase in required competencies also represents typical emerging market entry barriers, where organizations must develop entirely new capabilities to operate effectively in unfamiliar institutional environments.

Table 14 reveals significant differences in how expert groups assess current competency importance. For Sector-specific Domain competencies, DeFi Industry Practitioners rate these as significantly more important today (median = 4) than Course Participants (median = 2, H = 15.01, p = 0.001). Strategic competencies show a similar pattern, with DeFi Industry Practitioners valuing them more highly (median = 3) than Course Participants (median = 2, H = 12.73, p = 0.002). No significant experience-based differences emerge for current competencies. Looking ahead to 2034, Strategic competencies show marginal overall group differences (H = 6.66, p = 0.036), though post-hoc tests do not identify specific significant pairs. Notably, the median importance ratings for 2034 are consistently higher across all competency categories, confirming the universal expectation of increased DeFi literacy requirements.

From a NIE perspective, the current divergence in competency valuations reflects different positions within the institutional landscape. DeFi Industry Practitioners' significantly higher ratings for sector-specific and strategic competencies suggest they are actively building new institutional arrangements and thus immediately require these capabilities. Course Participants, primarily from traditional finance and regulatory bodies, may not yet face the same institutional pressures to develop DeFi-specific competencies, albeit internal training programs are underway within regulators around the globe. The convergence of ratings for 2034 indicates again expected institutional isomorphism; as DeFi becomes integrated into mainstream finance, all actors will require similar competency profiles.

The Dynamic Capabilities framework showcases why strategic competencies emerge as particularly critical. These competencies that include strategic vision, innovation management, and the ability to participate in DAOs does represent higher-order dynamic capabilities that enable organizations to sense opportunities, seize advantages, and transform operations in response to industry evolution over time. The universal recognition that all competencies will increase in importance by 2034 aligns with DCT's emphasis on continuous learning and adaptation. Organizations must develop not just DeFi sector-specific knowledge but comprehensive capabilities, also spanning technological, strategic, ethical, and interactional dimensions.

These insights underscore the capability gap faced by the financial services industry. The significant increase in importance across all competency categories suggests that current DeFi literacy levels are insufficient for the anticipated future. Organizations should prioritize comprehensive talent development programs that address multiple competency dimensions. The particularly sharp increases expected for strategic and DeFi sector-specific competencies indicate these should be immediate priorities. Furthermore, the current divergence between DeFi Industry Practitioners and other groups suggests that traditional financial institutions may need to accelerate their capability development to avoid being left behind.

Findings related to competencies extend traditional financial literacy to organizational contexts. While Lusardi and Mitchell (2014) identify numeracy, inflation understanding, and risk comprehension as core individual competencies, DeFi literacy requires six specific organizational dimensions.

**Table 14.** DeFi Competencies

**Panel A.** Statistical Analysis

| Variable | Expert Groups H | Expert Groups p-value | Expert Groups Significant Pairs | Experience Levels H | Experience Levels p-value | Experience Levels Significant Pairs |
|---|---|---|---|---|---|---|
| **Current Importance** | | | | | | |
| Sector-specific Domain | 15.01 | 0.001** | IP > CP | 0.90 | 0.826 | - |
| Strategic | 12.73 | 0.002** | IP > CP | 1.32 | 0.725 | - |
| Technological | 2.61 | 0.272 | - | 0.63 | 0.888 | - |
| Cognitive | 5.86 | 0.053 | - | 2.29 | 0.515 | - |
| Interactional | 0.55 | 0.759 | - | 2.07 | 0.558 | - |
| Ethical and Societal | 2.77 | 0.251 | - | 6.46 | 0.091 | - |
| Other | 1.81 | 0.405 | - | 2.81 | 0.421 | - |
| **Importance by 2034** | | | | | | |
| Sector-specific Domain | 2.80 | 0.246 | - | 1.81 | 0.614 | - |
| Strategic | 6.66 | 0.036* | - | 2.47 | 0.480 | - |
| Technological | 0.33 | 0.848 | - | 0.02 | 0.999 | - |
| Cognitive | 0.09 | 0.956 | - | 4.67 | 0.198 | - |
| Interactional | 0.62 | 0.732 | - | 2.12 | 0.549 | - |
| Ethical and Societal | 0.24 | 0.888 | - | 1.09 | 0.779 | - |
| Other | 0.80 | 0.669 | - | 4.05 | 0.256 | - |

**Panel B.** Group Medians (IQR)

| Variable | CP | AR | IP | 1-5 yrs | 5-10 yrs | 10-20 yrs | 20+ yrs |
|---|---|---|---|---|---|---|---|
| **Current Importance** | | | | | | | |
| Sector-specific Domain | 2.0 (1.0) | 3.0 (1.0) | 4.0 (1.0) | 3.0 (2.0) | 3.0 (2.0) | 3.0 (2.0) | 3.0 (2.0) |
| Strategic | 2.0 (1.0) | 3.0 (2.0) | 3.0 (1.0) | 3.0 (2.0) | 3.0 (1.0) | 3.0 (2.0) | 3.0 (2.0) |
| Technological | 3.0 (2.0) | 3.0 (0.5) | 3.0 (2.0) | 3.0 (2.0) | 3.0 (1.8) | 3.0 (2.0) | 3.0 (2.0) |
| Cognitive | 2.0 (1.0) | 3.0 (1.2) | 3.0 (1.0) | 3.0 (1.0) | 2.0 (2.0) | 2.5 (1.0) | 2.5 (1.0) |

| Variable | CP | AR | IP | 1-5 yrs | 5-10 yrs | 10-20 yrs | 20+ yrs |
|---|---|---|---|---|---|---|---|
| Interactional | 2.0 (1.8) | 2.5 (1.0) | 2.0 (1.0) | 2.0 (1.0) | 2.0 (2.0) | 2.0 (1.2) | 2.5 (1.0) |
| Ethical and Societal | 2.5 (2.0) | 3.0 (1.0) | 3.0 (2.0) | 3.0 (2.0) | 2.0 (2.0) | 3.0 (2.0) | 3.0 (2.0) |
| Other | 1.0 (2.0) | 1.0 (2.0) | 2.0 (2.5) | 1.0 (2.0) | 0.5 (2.0) | 1.0 (2.0) | 2.0 (2.8) |
| **Importance by 2034** | | | | | | | |
| Sector-specific Domain | 3.0 (1.0) | 3.0 (1.0) | 4.0 (1.0) | 3.0 (1.0) | 3.5 (1.8) | 4.0 (1.0) | 4.0 (1.0) |
| Strategic | 3.0 (1.0) | 3.0 (2.0) | 4.0 (1.0) | 4.0 (1.0) | 3.0 (1.8) | 4.0 (1.0) | 4.0 (1.0) |
| Technological | 4.0 (1.0) | 3.5 (1.2) | 3.0 (1.0) | 4.0 (1.0) | 3.5 (1.0) | 3.5 (1.0) | 3.0 (1.0) |
| Cognitive | 3.0 (1.8) | 3.0 (1.2) | 3.0 (1.0) | 3.0 (2.0) | 3.0 (1.0) | 3.0 (2.0) | 3.0 (1.0) |
| Interactional | 3.0 (1.8) | 3.0 (2.0) | 3.0 (1.0) | 3.0 (2.0) | 3.0 (1.8) | 3.0 (1.2) | 3.5 (1.0) |
| Ethical and Societal | 3.5 (1.8) | 3.5 (1.2) | 4.0 (1.0) | 3.0 (2.0) | 3.0 (1.8) | 4.0 (1.0) | 3.5 (1.0) |
| Other | 1.0 (3.0) | 1.5 (3.0) | 2.0 (3.0) | 1.0 (3.0) | 0.5 (2.8) | 1.5 (3.0) | 3.0 (2.8) |

Note: ** p < .01, * p < .05. CP = DeFi Course Participants, IP = DeFi Industry Practitioners, AR = DeFi-focused Academic Researchers. IQR = Interquartile Range shown in parentheses. A dash (-) in the Significant Pairs column indicates either no significant differences were found in the overall test or post-hoc pairwise comparisons did not reach significance after Bonferroni correction.

*Regional Perspectives*

Next I examine geographic variations by categorizing respondents into Asia-Pacific (n=67) and Western markets (n=38) based on headquarter location. I acknowledge that headquarters serves as an imperfect proxy for regional perspectives, particularly for multinational institutions.

Western market respondents demonstrate significantly higher expectations for DeFi adoption by 2034 (p=0.013) and more strongly believe traditional finance will embrace DeFi (p=0.001, r=0.32). They anticipate greater platform roles (p=0.003), expect highly regulated DeFi scenarios (p=0.012), and foresee stronger risk management impact (p=0.046). Western respondents also rate current strategic and DeFi sector-specific competencies as more important (p=0.011 and p=0.031 respectively). Regional risk concerns diverge notably. Western markets express greater concern about environmental impact (p=0.006), loss of human touch (p=0.009), and job displacement (p=0.014). Asia-Pacific respondents show higher concern only for funding constraints (p=0.038).

These regional differences align with institutional theory predictions of varying adoption patterns based on regulatory environments and market structures. Western optimism about traditional finance embracing DeFi (the strongest effect at r=0.32) again suggests expectations of institutional convergence rather than disruption.

From a Dynamic Capabilities perspective, Western respondents' higher valuation of current competencies may reflect their organizations' more advanced development of DeFi-related sensing capabilities. The insights presented indicate DeFi adoption may follow different regional trajectories, with Western markets potentially leading integration efforts while Asia-Pacific markets navigate resource allocation challenges. Table 15 presents Mann-Whitney U test results.

Despite Singapore's concentration in my sample, the regional differences offer valuable preliminary insights. Paradoxically, while Nguyen and Nguyen (2024) document high current DeFi adoption across Asian markets (Vietnam, Thailand, and China among top global adopters) my survey reveals more conservative future expectations in the Asia-Pacific region. This divergence between current adoption and future expectations suggests different underlying dynamics.

The regulatory landscape presents a complex picture that does not fully explain these regional differences. While the EU's MiCA regulation excludes DeFi (Schuler et al., 2024), Western respondents remain optimistic. Perhaps they anticipate future regulatory frameworks. Conversely, despite Singapore's Project Guardian actively experimenting with DeFi through regulatory sandboxes (Lim et al., 2023), Asia-Pacific respondents maintain more cautious outlooks. This may suggests that institutional sentiment about DeFi's future may be shaped by factors beyond regulatory clarity alone. It could include market maturity, prior experience with crypto volatility, and different risk appetites between retail adopters and institutional players.

**Table 15.** Regional Differences in DeFi Expectations. Shown are Mann-Whitney U test results comparing ratings of individual survey items between Asia-Pacific (n=67) and Western Markets (n=38) respondents. Each row represents a statement or scenario that experts rated.

| Variable | U-statistic | p-value | Asia-Pacific Median (IQR) | Western Markets Median (IQR) |
|---|---|---|---|---|
| TradFi Embraces DeFi | 781.0 | 0.001** | 3.0 (1.5) | 4.0 (1.0) |
| DeFi Platform Role | 843.0 | 0.003** | 3.0 (1.0) | 3.0 (1.0) |
| Impact on Environment | 1668.5 | 0.006** | 2.0 (2.0) | 2.0 (1.0) |
| Loss of Human Touch | 1649.5 | 0.009** | 2.0 (2.0) | 2.0 (1.0) |
| Strategic Competencies (Today) | 907.0 | 0.011* | 2.0 (2.0) | 3.0 (1.0) |
| Highly Regulated DeFi Worlds | 913.5 | 0.012* | 3.0 (2.0) | 3.0 (1.0) |
| Adoption 2034 | 922.5 | 0.013* | 2.0 (1.0) | 3.0 (1.0) |
| Job Losses/Upskilling | 1630.5 | 0.014* | 2.0 (3.0) | 2.0 (1.0) |
| Sector-specific Domain (Today) | 963.0 | 0.031* | 3.0 (2.0) | 3.0 (1.0) |
| Lack of Funding/Budget | 1570.5 | 0.038* | 3.0 (1.0) | 2.0 (2.0) |

*Note: ** p < .01, * p < .05. IQR = Interquartile Range. Four observations classified as "Other" excluded.*

*Post-Survey Developments: Validation of Expert Predictions*

In this last sub-section I compare expert elicitation results with recent developments in the form of SEC statements and speeches. SEC statements support rising DeFi adoption. Chairman Paul S. Atkins' June 9, 2025, remarks highlight DeFi's potential to enhance liquidity and efficiency, aligning with the survey's high adoption prediction. The SEC's May 29, 2025, clarification on staking not being a security supports the "Highly Regulated DeFi" scenario. New ETFs, like the Rex-Osprey ETH + Staking ETF, indicate traditional finance integrating DeFi, matching the "TradFi Embraces DeFi" scenario. Regulatory challenges persist, as noted by Commissioner Caroline A. Crenshaw, aligning with survey concerns. The SEC's focus on investor protection supports the survey's emphasis on risk management, and its emphasis on transparency may indirectly improve customer service in times ahead. SEC engagement, including roundtables, suggests a need for the development of DeFi Literacy. A full comparison is provided in Table 17.

**Table 17.** Alignment of SEC Statements with expert elicitation. Note: Alignment status based on author assessment. SEC statements issued up until June 2025. Strong alignment indicates direct support for survey predictions; Partial alignment indicates indirect or incomplete support.

| Survey Prediction | SEC Statement Evidence | Alignment Status | Source |
|---|---|---|---|
| High DeFi adoption by 2034 | Atkins praises DeFi's potential and resilience | Strong alignment; early trends support increased adoption | https://www.sec.gov/newsroom/speeches-statements/atkins-remarks-defi-roundtable-060925 |
| DeFi platforms as niche players or regulated actors | SEC clarifies staking activities are not securities, but centralized DeFi entities face scrutiny | Partial alignment; regulatory clarity supports regulated actors, but niche roles less clear | https://www.sec.gov/newsroom/speeches-statements/peirce-remarks-defi-roundtable-060925, https://www.sec.gov/newsroom/speeches-statements/statement-certain-protocol-staking-activities-052925 |
| "Highly Regulated DeFi" scenario | SEC's Crypto Task Force and proposed "innovation exemption" indicate structured regulation | Strong alignment; regulatory framework is evolving | https://www.sec.gov/newsroom/speeches-statements/crenshaw-remarks-crypto-roundtable-060925, https://www.sec.gov/newsroom/speeches-statements/atkins-remarks-defi-roundtable-060925, https://www.sec.gov/newsroom/speeches-statements/statement-certain-protocol-staking-activities-052925 |
| "TradFi Embraces DeFi" scenario | New ETFs including staking and SEC | Strong alignment; collaboration suggests integration | https://www.sec.gov/ix?doc=/Archives/edgar/data/1771146/000199937125006 |

| | | | 935/osprey-485bpos_053025.htm, https://www.sec.gov/newsroom/speeches-statements/crenshaw-remarks-crypto-roundtable-060925 |
|---|---|---|---|
| Regulatory challenges as barriers | Ongoing legal inconsistencies and state actions against staking services | Strong alignment; regulatory uncertainty persists | https://www.sec.gov/newsroom/speeches-statements/crenshaw-remarks-crypto-roundtable-060925, https://www.sec.gov/newsroom/speeches-statements/crenshaw-statement-protocol-staking-052925 |
| Impact on business areas (e.g., risk management) | SEC focus on investor protection and compliance aligns with risk management needs | Strong alignment; addresses survey's emphasis on risk | https://www.sec.gov/newsroom/speeches-statements/crenshaw-remarks-crypto-roundtable-060925 |
| Customer service improvements | Transparency and investor protection may indirectly enhance customer experiences | Partial alignment; moderate impact expected | https://www.sec.gov/newsroom/speeches-statements/crenshaw-remarks-crypto-roundtable-060925 |
| Increased DeFi competencies by 2034 | SEC engagement suggests need for strategic and regulatory skills | Strong alignment; supports competency development | https://www.sec.gov/newsroom/speeches-statements/peirce-remarks-defi-roundtable-060925, https://www.sec.gov/newsroom/speeches-statements/atkins-remarks-defi-roundtable-060925 |

## 7   Conclusion

DeFi is not a technology to be implemented but an emerging market to be entered. This paper examines how 109 experts across three distinct groups, Traditional Finance Professionals, DeFi Industry Practitioners, and DeFi-focused Academic Researchers expect this market to reshape financial services by 2034. I use non-parametric statistical methods for analysis and interpret results through two theoretical lenses: NIE to understand how DeFi reshapes financial market structures and business functions, and DCT to explain both required organizational adaptations and expertise-based perception differences.

I uncover four key insights that reframe the DeFi debate. First, despite varying current views, all expert groups agree on 2034 adoption expectations: while no group rates current institutional adoption as at least high, 43% anticipate at least high adoption within a decade. Second, experts expect convergence over competition, with traditional finance embracing DeFi within regulatory frameworks emerging as the most likely scenario. Third, DeFi will transform back-office operations before customer interfaces, with 86% expecting major risk management changes and 80% anticipating operations transformation. Fourth, strategic competencies will eclipse sector-specific DeFi- and technological competencies.

For financial services executives, these insights prescribe clear action: begin building DeFi literacy now, but prioritize strategic transformation capabilities over mere technical DeFi and blockchain training. The capability gap cited by 68% of experts represents both risk and opportunity and institutions that develop comprehensive competencies today will capture tomorrow's benefits. The expert predictions gain credibility from recent SEC developments.

For policymakers and regulators, the expert consensus on a "highly regulated DeFi" scenario suggests proactive frameworks could channel innovation productively. Rather than restricting development, clear regulatory guidelines could accelerate responsible institutional adoption while protecting financial system stability.

For the DeFi industry professionals, the expected convergence with traditional finance implies pivoting from disruption narratives toward integration strategies. Success will require understanding institutional needs and regulatory requirements, not just technological innovation.

This study opens multiple research avenues. Longitudinal studies could track actual adoption against expert predictions to validate forecasting accuracy. Comparative case studies of institutions developing DeFi capabilities could surface best practices. The dominance of strategic- over DeFi sector specific and technical competencies warrants investigation, too. Does this pattern hold across other FinTech business models? The regional paradox between retail and institutional adoption also deserves deeper exploration to understand how market structures shape novel emerging market entry pathways.

While 109 experts provide rich insights, the Singapore concentration (58%) invites replication across other major financial centres. The ten-year horizon, though reasonable for institutional planning, may miss nearer-term developments or longer-term transformations.

Ultimately, this study reveals DeFi literacy as organizational market entry capability rather than DeFi sector-specific or technical proficiency. The slow pace of institutional change, often lamented in financial services, becomes an advantage when directed toward systematic capability development. As experts across all groups converge on the importance of strategic competencies by 2034, the message is clear: in the race toward DeFi integration, strategic vision outpaces pure technical knowledge. Success belongs not to the fastest implementers but to the deepest thinkers about institutional transformation.

**References**

Adamyk, B, Benson, V., Adamyk, O., Liashenko, O., 2025. Risk Management in DeFi: Analyses of the Innovative Tools and Platforms for Tracking DeFi Transactions. *Journal of Risk and Financial Management*. https://doi.org/10.3390/jrfm18010038

Argyris, C., Schön, D.A., 1978. Organizational Learning: A Theory of Action Perspective. *Textbook*. https://archive.org/details/organizationalle00chri

Aquilina, M., Frost, J., Schrimpf, A., 2024. Decentralized Finance (DeFi): A Functional Approach. *Journal of Financial Regulation*. https://doi.org/10.1093/jfr/fjad013

Auer, R., Haslhofer, B., Kitzler, S. et al., 2023. The technology of decentralized finance (DeFi). *Digital Finance*. https://doi.org/10.1007/s42521-023-00088-8

BCBS, 2022. Prudential treatment of cryptoasset exposures. *Basel Committee on Banking Supervision*. https://www.bis.org/bcbs/publ/d545.htm

Bojke L, Soares M, Claxton K. 2021. Good practice in structured expert elicitation: learning from the available guidance. *Book Chapter in Health Technology Assessment*. https://www.ncbi.nlm.nih.gov/books/NBK571059/

Bongaerts, D., Lambert, T., Liebau, D., Roosenboom, P., 2025. Vote Delegation in DeFi Governance. *Working paper*. https://papers.ssrn.com/sol3/papers.cfm?abstract_id=5177022

Bok, K., 2024. Decentralizing Finance: How DeFi, Digital Assets and Distributed Ledger Technology Are Transforming Finance. *Textbook*. https://kennethbok.com/book

Buterin, V., Illum, J., Nadler, M., Schaer, F., Soleimani, A., 2024. Blockchain privacy and regulatory compliance: Towards a practical equilibrium. *Blockchain: Research and Applications*. https://doi.org/10.1016/j.bcra.2023.100176

Capponi, A., Jia, R., 2021. The Adoption of Blockchain-Based Decentralized Exchanges. *Working paper*. https://arxiv.org/pdf/2103.08842

Catalini, C., de Gotari, A., Shar, N., 2021. Some simple Economics of Stablecoins. *Working paper*. https://papers.ssrn.com/sol3/papers.cfm?abstract_id=3985699

Daian, P., Goldfeder, S., Kell, T., Li, Y., Zhao, X., Bentov, I., Breidenbach, L., Juels, A., 2019. Flash Boys 2.0: Frontrunning, Transaction Reordering, and Consensus Instability in Decentralized Exchanges. *Working paper*. https://arxiv.org/abs/1904.05234

Diamond, D. W., Dybvig, P. H., 1983. Bank runs, deposit insurance, and liquidity. *Journal of Political Economy*. https://www.jstor.org/stable/1837095

DiMaggio, P.J. and Powell, W.W., 1983. The Iron Cage Revisited: Institutional Isomorphism and Collective Rationality in Organizational Fields. *American Sociological Review*. https://doi.org/10.2307/2095101

Dion, P., Galbraith, N., Sirag, E., 2020. Using expert elicitation to build long-term projection assumptions. *Book Chapter in Developments in Demographic Forecasting.* https://library.oapen.org/bitstream/handle/20.500.12657/42565/2020_Book_DevelopmentsInDemographicForec.pdf#page=51

Dong, X., Blankson, C., 2025. Learning process challenges and outcomes in the fintech domain. *International Journal of Bank Marketing.* https://doi.org/10.1108/IJBM-05-2024-0263

Feichtinger, R., Fritsch, R., Heimbach, L., Vonlanthen, Y., Wattenhofer, R., 2024. SoK: Attacks on DAOs. *Working paper.* https://arxiv.org/abs/2406.15071

Fernandes, D., Lynch, J.G., Netemeyer, R.G., 2014. Financial Literacy, Financial Education, and Downstream Financial Behaviors. *Management Science.* https://doi.org/10.1287/mnsc.2013.1849

Graham, J., Harvey, C., 2001. The theory and practice of corporate finance: evidence from the field. *Journal of Financial Economics.* https://www.sciencedirect.com/science/article/pii/S0304405X01000447

Harvey, C., Ramachandran, A., Santoro, J., 2021. DeFi and the Future of Finance. *Book.* https://www.amazon.sg/DeFi-Future-Finance-Campbell-Harvey/dp/1119836018

John, K., Kogan, L., Saleh, F., 2023. Smart Contracts and Decentralized Finance. *Annual Review of Financial Economics.* https://www.annualreviews.org/doi/abs/10.1146/annurev-financial-110921-022806

Jones, M., Luu, T., Samuel, B., 2024. Measuring Cryptocurrency Literacy. *Journal of Behavioral Finance.* https://doi.org/10.1080/15427560.2024.2303421

Khalil, M., Khawaja, K., Sarfraz, M., 2021. The adoption of blockchain technology in the financial sector during the era of fourth industrial revolution: a moderated mediated model. *Quality & Quantity: International Journal of Methodology.* https://link.springer.com/article/10.1007/s11135-021-01229-0

Kreppmeier, J., Laschinger, R., Steininger, B., Dorfleitner, G., 2023. Real estate security token offerings and the secondary market: Driven by crypto hype or fundamentals? *Journal of Banking and Finance.* https://www.sciencedirect.com/science/article/pii/S0378426623001450

Lambert, T., Liebau, D., Roosenboom, P., 2021. Security Token Offerings. *Small Business Economics.* https://link.springer.com/article/10.1007/s11187-021-00539-9

Lehar, A., Parlour, C., 2023: Decentralized Exchange: The Uniswap Automated Market Maker. *Journal of Finance.* https://papers.ssrn.com/sol3/papers.cfm?abstract_id=3905316

Leland, H. E., Pyle, D. H., 1977. Informational asymmetries, financial structure, and financial intermediation. *The Journal of Finance.* https://www.jstor.org/stable/2326770

Liebau, D., Oh, Sandy, 2024. Decentralized Autonomous Organizations: How Finance can Interact with Blockchain-based DAOs. *Textbook.* https://doi.org/10.1142/13918

Lim, A., Tng, D., Lam, N., Ong, C.S., Pek, V., Rice, T., Shrakami, T., Hanock, J., 2023. Project Guardian – Enabling Open and Interoperable Networks. *Monetary Authority of Singapore / bank of international settlements*. https://www.mas.gov.sg/-/media/mas-media-library/development/fintech/project-guardian/project-guardian-open-interoperable-network.pdf

Liu, J., Makarov, I., Schoar, A., 2023. Anatomy of a run: The Terra Luna Crash. *NBER Working Paper*. https://www.nber.org/system/files/working_papers/w31160/w31160.pdf

Lusardi, A., Mitchell, O.S., 2011. Financial literacy and retirement planning in the United States. *Journal of Pension Economics and Finance*. https://doi.org/10.1017/S147474721000018X

Lusardi, A., Mitchell, O. S., 2014. The economic importance of financial literacy: Theory and evidence. *Journal of Economic Literature*. https://www.aeaweb.org/articles?id=10.1257/jel.52.1.5

Lusardi, A., Mitchel, O., 2023. The Importance of Financial Literacy: Opening a New Field. *Journal of Economic Perspectives*. https://www.aeaweb.org/articles?id=10.1257/jep.37.4.137

Kassoul, M., Prat, J., Tovanovich, N., Weidenholzer, S., 2024. Contagion in Decentralized Lending Protocols: A Case Study of Compound. *ACM Digital Library*. https://dl.acm.org/doi/10.1145/3605768.3623544

Malinova, K., Park, A., 2024. Learning from DeFi: Would Automated Market Makers Improve Equity Trading? *Working paper*. https://papers.ssrn.com/sol3/papers.cfm?abstract_id=4531670

Makarov, I., Schoar, A., 2022. Cryptocurrencies and Decentralized Finance (DeFi). *bis working paper*. https://www.bis.org/publ/work1061.pdf

Nguyen, L., Nguyen, P., 2024. Determinants of cryptocurrency and decentralized finance adoption - A configurational exploration. *Technological Forecasting and Social Change*. https://doi.org/10.1016/j.techfore.2024.123242

Nonaka, I., 1994. A Dynamic Theory of Organizational Knowledge Creation. *Organization Science*. https://doi.org/10.1287/orsc.5.1.14

North, D. C., 1990. Institutions, institutional change and economic performance. *Textbook*. https://www.cambridge.org/core/books/institutions-institutional-change-and-economic-performance/AAE1E27DF8996E24C5DD07EB79BBA7E

Ostrom, E., 2005. Understanding institutional diversity. *Textbook*. https://press.princeton.edu/books/paperback/9780691122380/understanding-institutional-diversity?srsltid=AfmBOorCHac9GvwOvNmfewIdibaGj795MvQA_kt7fbw0CYX4E6p34I1U

Popović, J., 2024. Uncovering the Strategic Potential of Blockchain Technology Adoption: A Systematic Literature Review. *Strategic Change*. https://doi.org/10.1002/jsc.2620

Rivera, T., Saleh, F., Vandeweyer, Q., 2023. Equilibrium in a DeFi Lending Market. *Working Paper*. https://papers.ssrn.com/sol3/papers.cfm?abstract_id=4389890

Saheb, T., Mamaghani, F., 2021. Exploring the barriers and organizational values of blockchain adoption in the banking industry. *Journal of High Technology Management Research*. https://doi.org/10.1016/j.hitech.2021.100417

Schär, F., 2021. Decentralized finance: On blockchain- and smart contract-based financial markets. *Federal Reserve Bank of St. Louis Review*. https://www.stlouisfed.org/publications/review/2021/02/05/decentralized-finance-on-blockchain-and-smart-contract-based-financial-markets

Teece, D. J. (2007). Explicating dynamic capabilities: The nature and microfoundations of (sustainable) enterprise performance. *Strategic Management Journal*. https://www.jstor.org/stable/20141992

Teece, D. J., Pisano, G., Shuen, A. (1997). Dynamic capabilities and strategic management. *Strategic Management Journal*. https://www.jstor.org/stable/3088148

Williamson, O. E., 1979. Transaction-cost economics: The governance of contractual relations. *The Journal of Law and Economics*. https://www.journals.uchicago.edu/doi/10.1086/466942

Williamson, O. E., 1985. The economic institutions of capitalism. *Textbook*. https://papers.ssrn.com/sol3/papers.cfm?abstract_id=1496720

Schuler, K., Cloots, A., Schaer, F., 2024. DeFi, CeFi, and the Regulatory Endgame. *Journal of Financial Regulation*. https://academic.oup.com/jfr/article/10/2/213/7606986

Zetsche, D.W., Buckley, R.P., Arner, D., 2020. Decentralized Finance. *Journal of Financial Regulation*. https://doi.org/10.1093/jfr/fjaa010

**Appendix 1 – Total Value Locked**

The path towards DeFi adoption, that is a wide use of DeFi primitives in financial services, is likely to encounter obstacles. A look at the total value locked or TVL tells a tale of volatility. TVL measures the total USD value of assets committed to DeFi protocols, offering a proxy for the ecosystem's size at a given time. The sharp increase in TVL in 2020, from circa USD 5 billion to circa USD 190 billion just a few months later, showcases how rapidly the adoption of DeFi progressed. But the sudden decline of TVL to circa USD 40 billion following the Terra LUNA crash, well described in Liu et al. (2023), in mid-2022 also highlights the fragility of the ecosystem. The losses associated with LUNA raised concerns about future potential systemic risks posed by DeFi, and international bodies like the Financial Stability Board (FSB) and regulators like the Bank for International Settlements (BIS) started investigating the phenomenon.[4] Though slower, the recent upward trend in TVL points to continued interest and perhaps a more sustainable growth phase. Singapore's Monetary Authority (MAS) was amongst the first globally well-respected regulators to investigate DeFi initiating Project Guardian in May of 2022.[5]

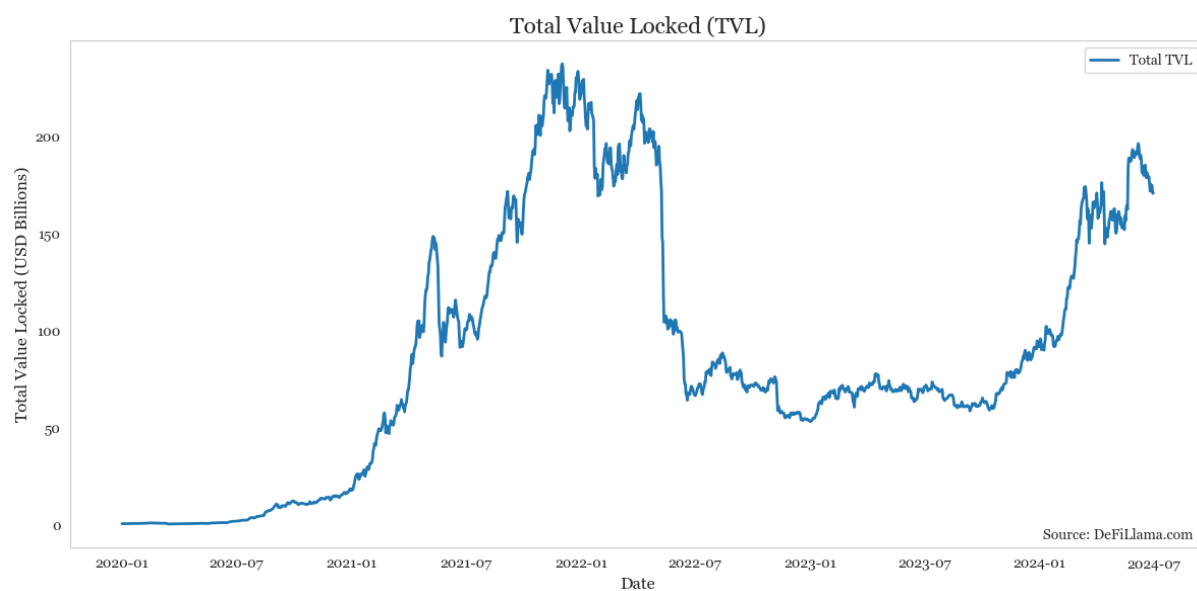

**Figure 1A**. Total Value Locked (TVL) in Decentralized Finance (DeFi) Protocols for the period from 1st Jan 2020 to 30th Jun 2024**.** This figure shows the total value locked (TVL) in DeFi protocols, data was sourced from DeFiLlama.com and excludes liquid staking and double-counted TVL according to the website's API endpoint description.

---

[4] Source: The Financial Stability Risks of Decentralised Finance, https://www.fsb.org/uploads/P160223.pdf (last accessed 28 Sep 2024)
[5] Source: https://www.straitstimes.com/tech/mas-launches-blockchain-project-to-study-decentralised-finance-potential-and-how-to-regulate-it (last accessed 15 Oct 2024)